\documentclass{emulateapj}
\usepackage{apjfonts}
\usepackage{natbib}
\usepackage{lscape}

\shorttitle{Radio AGN Lifetime}
\shortauthors{Bird, Martini, \& Kaiser}

\newcommand{\eg}{{\rm e.g.}}
\newcommand{\ie}{{\rm i.e.}}
\newcommand{\kms}{km s$^{-1}$}
\newcommand{\kgcm}{kg cm$^{-3}$}
\newcommand{\kgm}{kg m$^{-3}$}
\newcommand{\Tmax}{t$_{max}$}
\newcommand{\Qsad}{$Q_{S}$}
\newcommand{\Qblu}{$Q_{B}$}
\newcommand{\X}[1]{\ensuremath{\mathcal{X}_{#1}}}
\newcommand{\Rt}{\ensuremath{R_T}}
\newcommand{\LH}[1]{\ensuremath{\mathcal{L}_{#1}}}
\newcommand{\rfrac}{$r_{frac}$}
\newcommand{\rfraca}{$r_{frac,1}$}
\newcommand{\rfracb}{$r_{frac,2}$}
\newcommand{\rfracc}{$r_{frac,3}$}

\begin{document}

\title{The Lifetime of FRIIs in Groups and Clusters: Implications for Radio-Mode Feedback}

\author{Jonathan Bird, Paul Martini}
\affil{Department of Astronomy, The Ohio State University,
140 West 18th Avenue, Columbus, OH 43210, bird@astronomy.ohio-state.edu, martini@astronomy.ohio-state.edu}

\author{Christian Kaiser}
\affil{School of Physics and Astronomy, University of Southampton,
Southampton, SO17 1BJ U.K., crk@phys.soton.ac.uk}

\slugcomment{Accepted for publication in ApJ.}
\begin{abstract}
We determine the maximum lifetime \Tmax\ of 52 FRII radio sources found in 26 central group galaxies from cross correlation of the Berlind SDSS group catalog with the VLA FIRST survey. Mock catalogs of FRII sources were produced to match the selection criteria of FIRST and the redshift distribution of our parent sample, while an analytical model was used to calculate source sizes and luminosities. The maximum lifetime of FRII sources was then determined via a comparison of the observed and model projected length distributions. We estimate the average FRII lifetime is $1.5\times10^7$ years and the duty cycle is $\sim8\times10^8$ years. Degeneracies between \Tmax\ and the model parameters: jet power distribution, axial ratio, energy injection index, and ambient density introduce at most a factor of two uncertainty in our lifetime estimate.  In addition, we calculate the radio active galactic nuclei (AGN) fraction in central group galaxies as a function of several group and host galaxy properties. The lifetime of radio sources recorded here is consistent with the quasar lifetime, even though these FRIIs have substantially sub-Eddington accretion. These results suggest a fiducial time frame for energy injection from AGN in feedback models. If the morphology of a given extended radio source is set by large-scale environment, while the lifetime is determined by the details of the accretion physics, this FRII lifetime is relevant for all extended radio sources.
\end{abstract}


\section{Introduction}\label{sec:intro}

The interaction between active galactic nuclei (AGN) and their environment has been increasingly studied as a plausible explanation of several long standing problems in our understanding of galaxy formation and evolution. Hierarchal structure formation, first formalized by \citet{White78}, predicts a halo mass function that does not match the shape of the observed galaxy luminosity function (LF). Simulations predict an overabundance of massive and dwarf galaxies when compared to the observable universe \citep{White91, Benson03}. Another conflict between observation and theory is that galaxies are primarily red, early types with little star formation or actively star forming blue, late-types \citep[\eg][]{Blanton03, Kauffmann03, Kauffmann04}. This bimodality requires abrupt truncation of star formation. Most conspicuously, the temperature distribution of the intergalactic medium (IGM) of some clusters is at odds with simple radiative cooling arguments \citep{Peterson01, Hicks01}. The explanation of all of these problems favors regulation of galaxy evolution by an unknown mechanism of AGN `feedback'.

The strongest case for the existence of AGN feedback can be made in the cores of galaxy clusters and groups. Clusters have long been known as strong X-ray sources \citep{Byram66, Gursky71}, emitting thermal bremsstrahlung radiation from their hot gas \citep[first proposed and confirmed by][respectively]{Felten66, Mitchell76}. Calculations of the gas cooling time in some clusters showed that it was less than the Hubble time \citep{Lea73, Silk76}. Subsequently, \citet{Cowie77} and \citet{Fabian77} developed steady-state ``cooling flow'' models in which the gas of these systems condenses towards the center of the potential well. Two major problems with the cooling flow scenario have surfaced: (1) There have been no observations of the large star formation rate necessary for the center of clusters to be mass sinks of the condensing gas \citep{Fabian91} and (2) High resolution X-ray observations confirm that cluster centers have far less gas below $1$ keV than predicted \citep{Peterson01, Allen01, Hicks02, Fabian03}. As evidence mounts that gas in these systems do not exhibit strong cooling flows, they have recently been renamed ``cool core'' clusters \citep{Molendi01, Donahue04}. 
 
While cooling flows have been discounted, galaxies exist, so baryons must cool. Removing radiative cooling in the intracluster medium (ICM) produces the wrong slope of the X-ray luminosity-temperature relation ($L_X-T_X$) \citep{KaiserN86, Borgani01, Muanwong01}. To resolve these discrepancies, radiative cooling must be balanced, at least in part, by some heating source. Initially, feedback from supernovae was thought to heat the ICM, however, it is now assumed that only AGN could produce the enormous energy needed to combat cooling \citep[\eg][]{Valageas99, Wu00}. Early evidence for an AGN connection came from cross correlating cluster samples with radio sources \citep{Owen74, Bahcall74}. Subsequent study by  \citet{Burns90} found that more than $70\%$ of cluster dominant galaxies in cool core clusters are luminous radio galaxies.  This radio-X-ray correlation led to growing interest in the possibility of AGN feedback manifested as ICM heating via radio lobes \citep[\eg][]{Binney95, Churazov01, Soker01, Bruggen02, Begelman04}.

Spectacular support of AGN feedback on the ICM came from the observations of large X-ray cavities in the gas surrounding group and cluster cores spatially coincident with the radio lobes of jets \citep{Mcnamara00,Fabian00}. The prototype of radio mode feedback in clusters is now the set of extremely deep {\it Chandra} observations of Perseus, which show large cavities in the gas near the cluster core and evidence for shocks, ripples, and sound waves in the ICM \citep{Fabian03, Fabian06}. Observations of Cygnus A also exhibit clear interactions between the central radio source and the hot gas in this group of galaxies \citep[\eg][]{Wilson00}. ICM heating via radio mode feedback would eliminate the need for a central mass sink and create the ``entropy floors'' seen in clusters \citep{Valageas99, Nath02}, producing the observed $L_X-T_X$ slope \citep{Evrard91, KaiserN91}. Most convincingly, the amount of energy necessary to heat the ICM is consistent with the energy needed to physically create the cavities in many systems \citep{Birzan04}. AGN feedback probably occurs over a broad range of energies and efficiencies: the sizes of a few radio cavities suggest that their creation is not sufficient to completely balance radiative cooling \citep{Pope07}. The actual heating mechanism employed by radio lobes is still under intense investigation \citep[\eg][]{Ruszkowski04, Reynolds05}.

The timescale of AGN feedback is unknown yet fundamental to understanding its underlying physics and to model ICM evolution and the accretion history of AGN. Several different methods have been introduced to estimate the AGN feedback lifetime via the radio emission of lobes or X-ray observations of the ICM. \citet{Omma04} assumed that radio cavities move at the sound speed of the ambient medium and calculate their age using their projected distance from the galaxy core. The time necessary for cavities to buoyantly rise to their present position can provide an estimate of their lifetime as well \citep{Churazov01}.  \citet{Mcnamara00} and \citet{Nulsen02} determined cavity age using the time required to refill the volume of ICM displaced by the rising bubbles. Cavity ages between $\sim 10^6$ and $\sim10^8$ years have been recorded using these techniques \citep[\eg][]{Birzan04}. However, some of these age estimates apply to the current AGN age, or the time since the AGN turned off. The analysis herein estimates the average total lifetime of FRII sources and consequently the timescale of their AGN energy injection. This energetic AGN lifetime is critical to determining the physical processes of AGN feedback and whether radiative cooling can be balanced by said mechanisms. 

Radio loud AGN are not in every cool core cluster; thus, if radio-mode feedback is viable, it must be episodic. \citet{Binney95} were among the first to describe evolutionary models of the ICM. Radiative loses and gravitational forces promote gas infall towards the potential well and  provide fuel for the AGN. The AGN then forms jets which in turn lead to the creation of lobes that heat the ICM and cause a net outflow of gas. Once this fuel supply is removed, the AGN can no longer energetically sustain the jet and the newly created cavities then rise buoyantly through the ICM. The strongest evidence for episodic heating has been the discovery of ``ghost'' cavities in the outer regions of some clusters coincident with weaker radio emission no longer directly associated with the AGN \citep{Mcnamara01, Johnstone02, Mazzotta02}.

The preponderance of evidence supporting a connection between galaxy evolution and AGN has led to the assumption that black hole growth regulates star formation \citep{Silk98, Fabian99, King03, Weinmann06}. AGN feedback provides a natural regulation mechanism: the energy injected into the interstellar medium (ISM) prevents large scale star formation and provides an upper limit to the stellar mass of the galaxy, contributing to galaxy downsizing \citep[\eg][]{Cowie96, Scannapieco05}. To better understand the consequences of AGN feedback, hydrodynamic models have begun detailing the heating affect of radio lobes in the group and cluster environment \citep[\eg][]{Sijacki06}. Hierarchal structure formation models now routinely invoke AGN feedback to obtain the observed shape of the galaxy luminosity function and bimodal galaxy distribution \citep[\eg][]{Granato04, Springel05a,Springel05b,Bower06, Croton06, Hopkins06}. In addition to the nature of the physical process that couples the radio source to the ICM, the two main sources of uncertainty in the aforementioned models are the efficiency of energy injection and again, the timescales over which the process occurs. We focus on the latter unknown to provide context for the consequences of AGN feedback and its role in galaxy evolution.

Some new insights into the details of the dynamics of radio lobes and how they transfer energy to the surrounding medium can be gained from high-resolution observations of individual sources. Multicomponent radio sources are usually divided into one of two morphological classes: Fanaroff and Riley Class I and II \citep{Fanaroff74}. FRI sources are dominated by emission towards the nucleus with a gradient towards fainter and more diffuse emission as one moves radially outward. FRIIs are limb brightened sources, with well defined areas of brightest emission (``hotspots'') in their outermost regions. In addition to their morphological distinction, FRIIs are generally more luminous than FRIs \citep{Fanaroff74} while the most powerful FRIIs usually show a strong AGN component in their spectra indicative of obscured quasars \citep{Barthel89, Hes93}. FR class may also be a function of environment as the majority of extended radio sources in clusters are FRIs \citep{Prestage88, Hill91}. FRI sources may be FRIIs that encountered high density environments as they arise from turbulent disruption of their lobes and entrainment of ambient gas. Two morphological groups within the FRI distinction, Narrow Angle Tails (NATs) and Wide Angle Tails (WATs), are the result of radio source interaction with the ICM in dense cluster and group environments, either displaced due to galactic motion relative to the ambient medium \citep{Owen76} or bulk motions in the ICM \citep{Burns94}. Both FRI and FRII sources have been observed interacting with the surrounding medium \citep[\eg][]{Blanton00, Fabian03, Wilson03}. 

The well defined structure of FRIIs is an astrophysical analog of a supersonic, collimated flow into ambient gas \citep{DeYoung02}. Long identified as the collision site of the energetic jet particles and the surrounding medium, the hotspot position of FRII sources has been used as a standard measuring stick of the spatial extent of these powerful radio sources and can potentially reveal source age. Asymmetries in the length of classic double lobed sources due to light travel time have been used to estimate their expansion speed \citep{Longair79, Arshakian04}. High resolution measurements of these geometric asymmetries estimate the hot spot advance speed to be $\sim 0.02c$ throughout the jet's lifetime \citep{Scheuer95}. The radiative properties of radio lobes can provide estimates of the hotspot advance speed and source age as well. Radiating particles advect away from the hotspot and fill the cocoon, losing energy and softening the radio spectrum of the source as a function of age \citep{Alexander87a, Alexander87b}. Detailed observations of Cygnus A show that the spectral index is a strong function of radial distance from the AGN \citep{Carilli91}; this is seen in other FRII sources as well. Recent spectral age estimates of FRIIs are between $10^6$ and $10^7$ years (O'Dea 2007, in prep.). 

Shared characteristics of FRII sources led to the idea of self-similar growth \citep{Falle91}. In addition to similar advance speeds, similar axial ratios (\Rt, the ratio of a lobe's length to width) were recorded for FRII sources over a wide range of linear sizes \citep{Leahy84, Leahy89, Subrahmanyan96}. According to  the basic model for FRII growth, an AGN gives rise to two jets that propagate in opposite directions \citep{Blandford74,Scheuer74}. The energetic jet terminates in a shock, forming the hotspot, and drives a bow shock into the ambient medium. This accepted foundation has been built on by a number of models that analytically describe the dynamics of FRII sources \citep[\eg][hereafter KA]{Kaiser97a} and their luminosity evolution \citep[\eg][hereafter  KDA, BRW, MK, respectively]{Kaiser97b, Blundell99, Manolakou02}. All three use the dimensional arguments of KA to determine the length of FRII sources as a function of age but differ on how energy is deposited in the radio lobes from the hotspot. Given the physical properties of the jet and its ambient environment, each model predicts the track of sources in the luminosity-size plane. We use the KDA model as the crucial link between the intrinsic and observable characteristics of FRIIs. While the physics of radio jets are not completely understood in detail, these models of FRII source evolution produce results consistent with the demographics of large radio surveys \citep{Barai06}. 

Demographic studies of radio mode feedback have been greatly assisted by the recent completion of large area optical and radio sky surveys. In particular, the Sloan Digital Sky Survey (SDSS), covering two fifths of the Northern Galactic sky down to an $r$-band limiting magnitude of $\sim 22.5$, has proved invaluable in determining where potential radio mode feedback systems (groups and clusters) are on the sky. Several systematic searches for clusters and groups using photometric \citep[\eg][]{Annis99, Kim02, Goto02, Lee04} and spectroscopic \citep[\eg][]{Goto05, Miller05, Merchan05} data have provided valuable catalogs for a plethora of research.  The Faint Images of the Radio Sky at Twenty centimeters \citep[][FIRST]{Becker95,White97} survey is the radio analog to SDSS. Surveying the SDSS area down to 1 mJy at 1.4 GHz, FIRST represents a substantial improvement in radio surveys with regards to sensitivity and angular resolution. Other radio surveys, especially the NRAO VLA Sky Survey \citep[][NVSS]{Condon98}, and FIRST constitute an expansive and detailed picture of the radio universe. The combination of these optical and radio resources provides unparalleled opportunity to investigate the radio properties of the richest systems in the universe. In addition to our FRII lifetime estimate, we present a demographic study of radio sources in low redshift central group galaxies.

One of the first investigations from these large area surveys has been the measurement of the radio galaxy LF and its deviations from the optical galaxy LF \citep{Sadler02, Mauch07}. Correlations between radio active AGN and host galaxy properties, including local density and luminosity, have strengthened the arguments for radio mode feedback \citep{Best05a, Best05b}. In addition, a higher fraction of radio loud AGN are found in galaxies at the center of clusters compared to similar galaxies not at the center of their cluster's potential well \citep{Best06a, Croft07}. Demographic studies have provided good evidence that radio mode feedback is a factor in galaxy evolution and even regulation.

In the present study we produce a sample of FRII radio sources associated with central group galaxies using optical data from the SDSS and radio maps from FIRST to measure the maximum lifetime of FRIIs. Our systematic search for and subsequent analysis of FRIIs measures their typical energetic lifetime-- a fundamental parameter necessary to ascertain the nature of AGN feedback and its ramifications. The properties of the host galaxy are assembled from our optical data while the source attributes are obtained from our radio data. We ultimately produce a collection of FRII radio sources and measure their projected lengths (Section~\ref{sec:sample}). Simulations of FRII sources with known properties are produced in a Monte Carlo scheme using the KDA model and compared to our observed sample in Section~\ref{sec:popgen}. In Section~\ref{sec:res} we present our results for the lifetime of FRII sources, while dependence upon group and host galaxy properties is discussed in Section~\ref{sec:res_clustering}. We summarize our findings in Section~\ref{sec:con}.

\section{Sample}\label{sec:sample}

\subsection{Optical Sample} \label{sec:optsample}
This study's primary sample is a subset of the group and cluster catalog created by \citet{Berlind06a} using the SDSS. Using a dedicated $2.5$ meter telescope at Apache Point Observatory, SDSS scans the sky in five photometric bandpasses simultaneously with its $120$ Megapixel camera \citep{Gunn98}. Except in the case of fiber collisions owing to very small angular separations on the sky, most galaxies in the survey are selected for spectroscopic follow up.
\citet{Berlind06a} use a redshift space friends of friends (FoF) algorithm to identify groups of galaxies within the {\sc sample14} galaxy set \citep{Blanton03}; itself a subset of the main SDSS galaxy sample \citep{Strauss02}. FoF identification schemes are based on the premise of recursively linking galaxies together within a specified linking volume around each galaxy. The specific details can be found in \citet{Berlind06a} and sources cited therein. FoF algorithms have several advantages: they produce unique groups of galaxies for a given linking volume, no group geometry (\eg\ spherical) need be assumed a priori, and groups are only supplemented with new members when the linking volume is increased.
The \citet{Berlind06a} group catalog is a volume limited sample of $57138$ galaxies complete down to M$_{r}=-19.9$ mag and spanning the redshift range $0.015<z<0.1$. Each group member's luminosity and spectroscopic redshift, in addition to the total group luminosity, are reported in the catalog.

\subsection{Radio Sample}\label{sec:radsample}
We used the FIRST \citep{Becker95} survey conducted at the VLA to identify radio sources in our group sample. FIRST, a radio survey conducted at 1.4 GHz, has an angular resolution of $\sim 5\arcsec$ and a typical RMS of $0.15$ mJy per pixel. The resulting catalogs have a source detection limit of 1 mJy and astrometry better than $1\arcsec$. The survey's area was designed to match that of SDSS. FIRST represents roughly an order of magnitude improvement in angular resolution and sensitivity over previous large area radio sky maps. These survey  properties allow us to identify extended radio sources associated with single galaxies even in a crowded group environment.

FIRST observations of FRII sources were complimented with data from the NRAO VLA Sky Survey (NVSS). Also carried out at 1.4 GHz, NVSS is complete down to 2.5 mJy and has an angular resolution of 45 arcseconds \citep{Condon98}. Due to NVSS's shorter baselines in comparison to FIRST, it is much less likely to resolve out any flux from sources.

\subsection{Cross-Matching of the Optical and Radio Catalogs}\label{sec:cross}
To create a manageable sample size, groups with total M$_{r} \le -22.0$ mag were included in this study. In addition, groups whose central galaxy's coordinates were not included in FIRST were excluded. The remaining sample is composed of $2020$ groups with mass greater than $\sim10^{13.5} h^{-1} M_{\odot}$ and a typical velocity dispersion range of $100 \le \sigma_v \le 700$ \kms. Central group galaxies, identified in \citet{Berlind06a} as the brightest group member, were cross-correlated with FIRST.

We note that the FIRST catalog reports radio ``components'' as opposed to whole sources \citep{White97}. As such, any FRII source would have two components \citep{deVries06}. We searched for multicomponent $(\ge 2)$ radio sources in the FIRST catalog within $100$ kpc of each central galaxy. Each candidate was then visually inspected for FRII morphological features (limb brightened double sources). FR morphology can be robustly detected when the source is $\ge10\arcsec$ (corresponding to the minimum FIRST component size plus the approximate half power beam width of FIRST) and we require this minimum length of any potential FRII.  We find 32 central galaxies whose associated radio sources meet this simple criteria. Any FRI interlopers are removed from the sample via the morphological distinction made by \citet{Fanaroff74}, \ie, if the ``ratio of the distance between the regions of highest brightness of opposite sides of the central galaxy or quasar to the total extent of the source'' was greater than 0.5, the source is an FRII \citep{Fanaroff74}. Of our 32 potential candidates, 26 extended radio sources pass this morphological cut. These galaxies and their 52 associated lobes comprise our sample and their properties are summarized in Table~\ref{table:Sample}. Thumbnails of the optical and radio emission of each galaxy in our sample are shown in Figure~\ref{fig:sample1}. We discuss determination of lobe and ``hotspot'' length in Section ~\ref{sec:meas}.

After the initial FR morphological distinction by \citet{Fanaroff74}, FR sources have historically been classified using an empirically seen break between the luminosity of the two classes \citep{Fanaroff74, Ledlow96}. \citet{Ledlow96} show the break is a function optical luminosity ($L_{rad} \propto L_{opt}^{1.8}$) but is less well-defined in clusters and at lower redshifts. The mean $M_r$ of the host galaxies in our sample is -21.82, corresponding to an FRI/FRII break luminosity of $L_{1.4 Ghz} \sim 4.7\times10^{24}$ W Hz$^{-1}$ at $z=0.1$ for our choice of cosmology (20 sources do not exceed this threshold in our sample, of which 9 are within a factor of $\sim3$).  We note that studies of FR sources have been dominated by high luminosity sources due to the relative lack of sensitivity in wide-area radio surveys until recent years. \citet{Blundell99} argue that morphology is a much more fundamental quantity than luminosity as similarly luminous sources will not have the same jet properties at all redshifts. Our sample is strictly composed of FRIIs in the morphological sense. Further, KA and KDA model the self similar evolution of limb brightened lobes with distinct working surfaces. Our sample meets this criteria and thus the KA and KDA models can be applied appropriately. 

\begin{deluxetable*}{lcccccccccccccc}
\tablewidth{0pt}
\tabletypesize{\tiny}
\tablecolumns{15}
\tablecaption{FRII Source Sample\label{table:Sample}}
\tablehead{
\colhead{SDSS ID} &
\colhead{$z$} &
\colhead{$M_{r,grp}$} &
\colhead{(g-r)$_{grp}$} &
\colhead{$M_{r,cen}$} &
\colhead{(g-r)$_{cen}$} &
\colhead{$F_{FIRST}$} &
\colhead{$F_{NVSS}$} &
\colhead{$P_{NVSS}$} &
\colhead{$l_{hs,1}$} &
\colhead{$l_{hs,2}$} &
\colhead{$l_{p,1}$} &
\colhead{$l_{p,2}$} &
\colhead{$l_{p,1}$} &
\colhead{$l_{p,2}$}  \\
\colhead{} &
\colhead{} &
\colhead{} &
\colhead{} &
\colhead{} &
\colhead{} &
\colhead{(mJy)} &
\colhead{(mJy)} &
\colhead{($10^{23}$ W Hz$^{-1}$)} &
\colhead{(kpc)} &
\colhead{(kpc)} &
\colhead{($\arcsec$)} &
\colhead{($\arcsec$)} &
\colhead{(kpc)} &
\colhead{(kpc)} \\
\colhead{(1)} &
\colhead{(2)} &
\colhead{(3)} &
\colhead{(4)} &
\colhead{(5)} &
\colhead{(6)} &
\colhead{(7)} &
\colhead{(8)} &
\colhead{(9)} &
\colhead{(10)} &
\colhead{(11)} &
\colhead{(12)} &
\colhead{(13)} &
\colhead{(14)} &
\colhead{(15)} \\
}

\startdata
J020217.2-010740.2&	0.042&	-23.40&	0.97&	-21.76&	1.04&	32.57	&40.4&	1.62&	19.9&	16.9&  33.9	& 26.9 &	27.7&	22.0\\
J073600.8+273926.0&	0.078&	-22.61&	1.16&	-21.57&	1.23&	11.97	&12.0&	1.75&	11.2&	6.3&   10.0	& 10.1 &	14.5&	14.7\\
J074535.7+335746.6&	0.062&	-23.38&	0.91&	-21.09&	0.98&	70.07	&82.9&	7.47&	13.7&	15.2&  19.3	& 24.7 &	22.7&	29.1\\
J075625.7+370329.6&	0.076&	-23.25&	0.94&	-21.61&	1.00&	60.67	&208.4&	28.80&	71.2&	47.2&  86.0	& 47.8 &	122.2&	67.9\\
J075828.1+374711.8&	0.040&	-22.81&	1.02&	-22.28&	1.05&	546.07	&2717.9&	98.70&	42.8&	46.5&  74.2	& 65.7 &	57.9&	51.3\\
J080113.2+344030.8&	0.081&	-23.11&	0.92&	-21.33&	1.00&	23.27	&44.4&	7.02&	21.0&	45.1&  23.6	& 40.3 &	35.5&	60.7\\
J081023.2+421625.8&	0.063&	-23.33&	0.97&	-22.06&	1.03&	8.47	&10.1&	0.94&	7.7&	6.0&   10.5	& 10.2 &	12.5&	12.2\\
J084632.4+293555.3&	0.069&	-23.57&	0.91&	-21.98&	1.04&	76.77	&89.4&	10.08&	18.2&	21.5&  24.5	& 25.1 &	31.9&	32.6\\
J084759.0+314708.3&	0.066&	-24.09&	0.96&	-22.62&	1.01&	323.27	&868.7&	89.25&	103.6&	99.8&  120.5	& 88.9 &	150.4&	111.0\\
J093058.7+034827.7&	0.087&	-23.56&	0.99&	-22.07&	1.00&	31.87	&57.5&	10.58&	38.8&	34.9&  37.7	& 34.8 &	60.5&	55.9\\
J094708.8+421125.6&	0.071&	-22.16&	0.98&	-21.59&	1.00&	26.37	&39.4&	4.72&	44.9&	37.4&  42.5	& 35.8 &	56.7&	47.8\\
J102204.6+445144.0&	0.081&	-22.04&	0.99&	-21.64&	1.00&	396.67	&385.8&	61.01&	19.3&	22.1&  26.5	& 24.4 &	39.9&	36.7\\
J104958.8+001920.2&	0.039&	-23.15&	0.88&	-21.65&	1.00&	18.27	&61.8&	2.13&	12.9&	11.3&  27.2	& 21.2 &	20.7&	16.1\\
J114506.5+533853.0&	0.068&	-22.49&	0.95&	-21.91&	1.02&	103.87	&147.8&	16.17&	67.3&	53.9&  62.6	& 50.0 &	80.3&	64.2\\
J115011.2+534321.0&	0.060&	-24.11&	0.84&	-21.49&	0.98&	71.97	&82.5&	6.94&	18.5&	17.0&  24.1	& 22.9 &	27.5&	26.2\\
J122718.3+085036.8&	0.087&	-24.27&	0.90&	-21.27&	0.95&	46.47	&47.7&	8.78&	19.9&	25.2&  19.3	& 22.7 &	31.0&	36.5\\
J130239.0+622939.6&	0.074&	-23.23&	0.99&	-22.07&	1.03&	287.07	&306.3&	40.02&	28.7&	23.2&  31.7	& 33.7 &	44.0&	46.7\\
J133151.8-025219.5&	0.085&	-22.24&	0.96&	-21.92&	1.00&	72.07	&84.6&	14.82&	21.8&	23.4&  22.1	& 24.5 &	34.8&	38.5\\
J135442.2+052856.0&	0.076&	-22.66&	0.87&	-21.98&	0.96&	37.97	&36.5&	5.04&	17.6&	23.2&  21.5	& 23.2 &	30.5&	33.0\\
J150315.1+360851.8&	0.072&	-23.03&	0.96&	-22.13&	0.99&	189.77	&402.5&	49.64&	60.7&	63.6&  60.2	& 59.7 &	81.4&	80.7\\
J154417.8+344146.6&	0.071&	-23.70&	0.81&	-21.92&	0.83&	119.57	&199.1&	23.85&	42.1&	49.4&  45.4	& 45.4 &	60.6&	60.6\\
J155721.4+544016.2&	0.046&	-22.57&	0.98&	-21.67&	0.99&	76.27	&96.2&	4.66&	26.2&	44.1&  46.8	& 58.0 &	41.7&	51.7\\
J160804.5+430948.4&	0.083&	-23.51&	0.95&	-21.84&	0.98&	62.77	&74.4&	12.39&	21.6&	21.2&  23.8	& 23.5 &	36.6&	36.2\\
J171329.0+640248.9&	0.077&	-23.46&	0.94&	-21.95&	1.05&	252.67	&265.7&	37.75&	23.8&	18.1&  21.0	& 19.3 &	30.2&	27.8\\
J215701.7-075022.5&	0.061&	-24.70&	0.78&	-22.05&	0.81&	366.57	&433.1&	37.74&	34.6&	33.4&  41.2	& 41.7 &	47.8&	48.4\\
J235958.8+004206.3&	0.080&	-23.54&	0.89&	-21.88&	0.91&	270.17	&411.1&	63.32&	124.7&	88.5&  95.6	& 85.6 &	142.3&	127.4\\
\enddata

\tablecomments{FRII sources in our sample in order of increasing right ascension. (1): SDSS identifier. (2): Galaxy redshift. (3): Total r-band absolute magnitude of the group. (4): (g-r) color of host group. (5): Absolute r-band magnitude of the host galaxy. (6): (g-r) color of the host galaxy. (7): Total source flux in mJy reported by FIRST. (8): Total source flux in mJy reported by NVSS. (9): Total source luminosity in $10^{23}$ W Hz$^{-1}$ computed from the NVSS flux. (10,11): The projected separation of the hotspot and central galaxy in kpc. (12,13): The angular size of the two lobes in arcseconds. (14,15): The projected length of the two lobes in kpc. Note that (10,12,14) will always correspond to the east hotspot and lobe, (11,13,15) to the west hotspot and lobe.}
\end{deluxetable*}

\begin{figure*}[tbp]
\figurenum{1} 
\includegraphics[width=\textwidth]{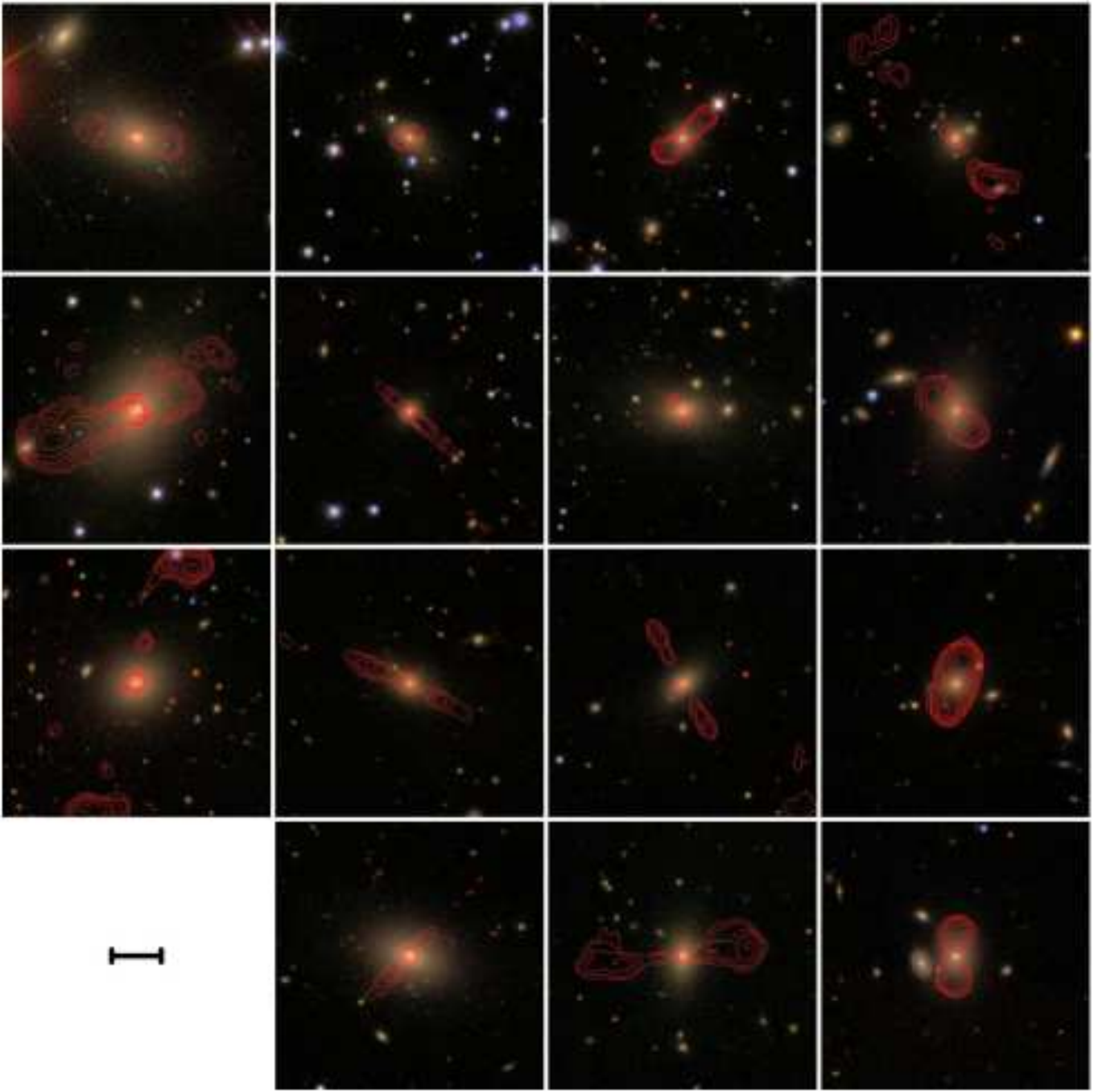}
\caption{\label{fig:sample1} The host galaxies with FRII radio sources in our sample. The SDSS r-band images are the grayscale while the contours are logarithmic radio flux from FIRST. Each image is $162\arcsec$ on a side. The scale bar in the bottom left is $30\arcsec$.}
\end{figure*}

\begin{figure*}[tbp]
\figurenum{1} 
\includegraphics[width=\textwidth]{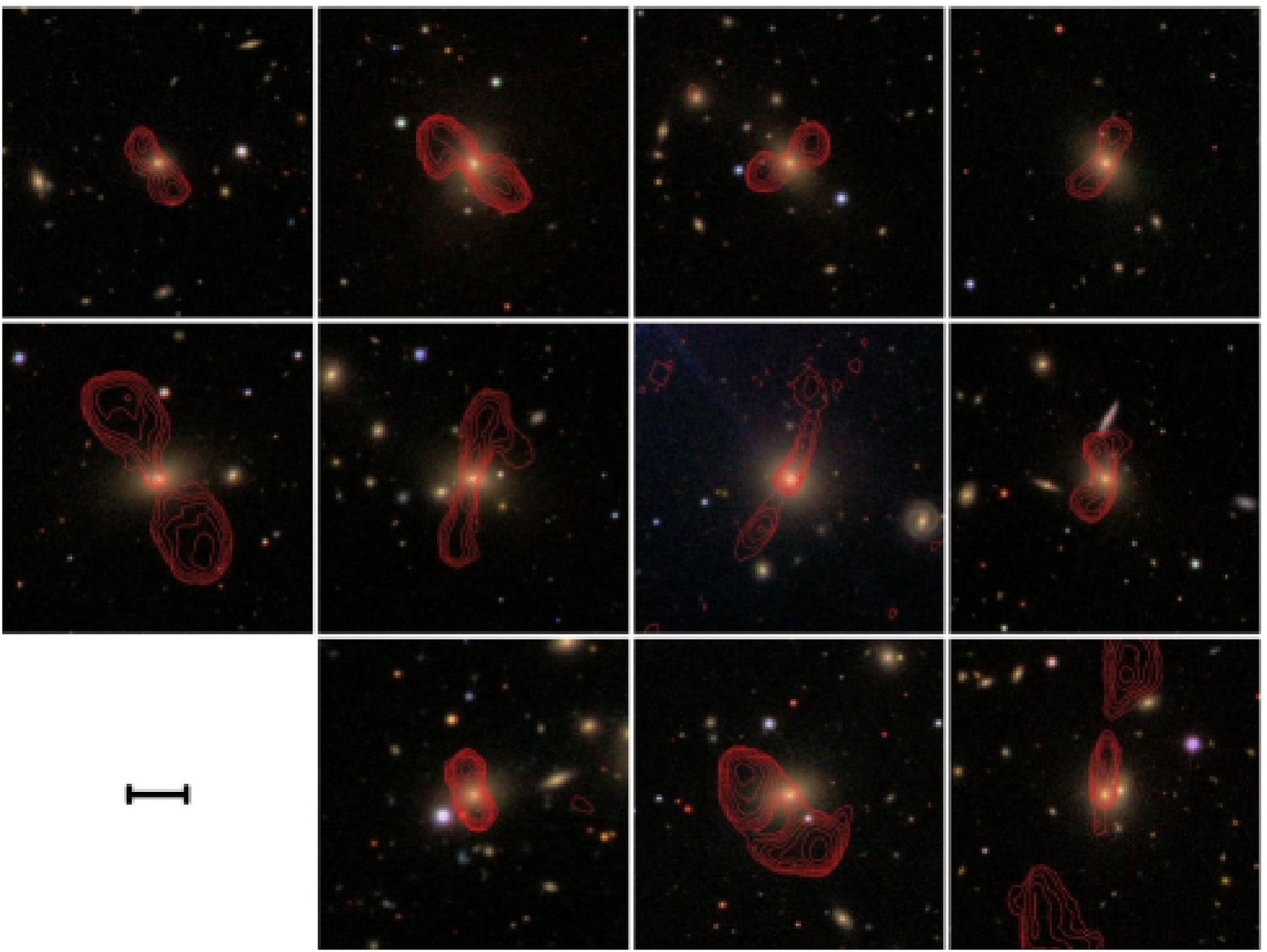}
\caption{\label{fig:sample1} continued.}
\end{figure*}

\subsection{Projected Source Length Measurement}\label{sec:meas}
Several methods have been employed in the literature to measure the projected length of radio lobes. In the past, a few studies investigated the ratio of projected lobe lengths of classic double lobed radio sources to determine their expansion velocities \citep{Longair79,Scheuer95}. \citet{Scheuer95} measured the projected lobe length as the distance between the ``second or third'' radio contour at the edge of the lobe and the core radio source, noting ``that any well-defined prescription, rigorously applied, will sometimes produce obvious nonsense.'' In more recent work, the length of the source is defined as the distance from the ``hotspot,'' or the termination shock of the jet, to the central radio source.

\begin{figure}[tbp]
\epsscale{1.15}
\figurenum{2} 
\plotone{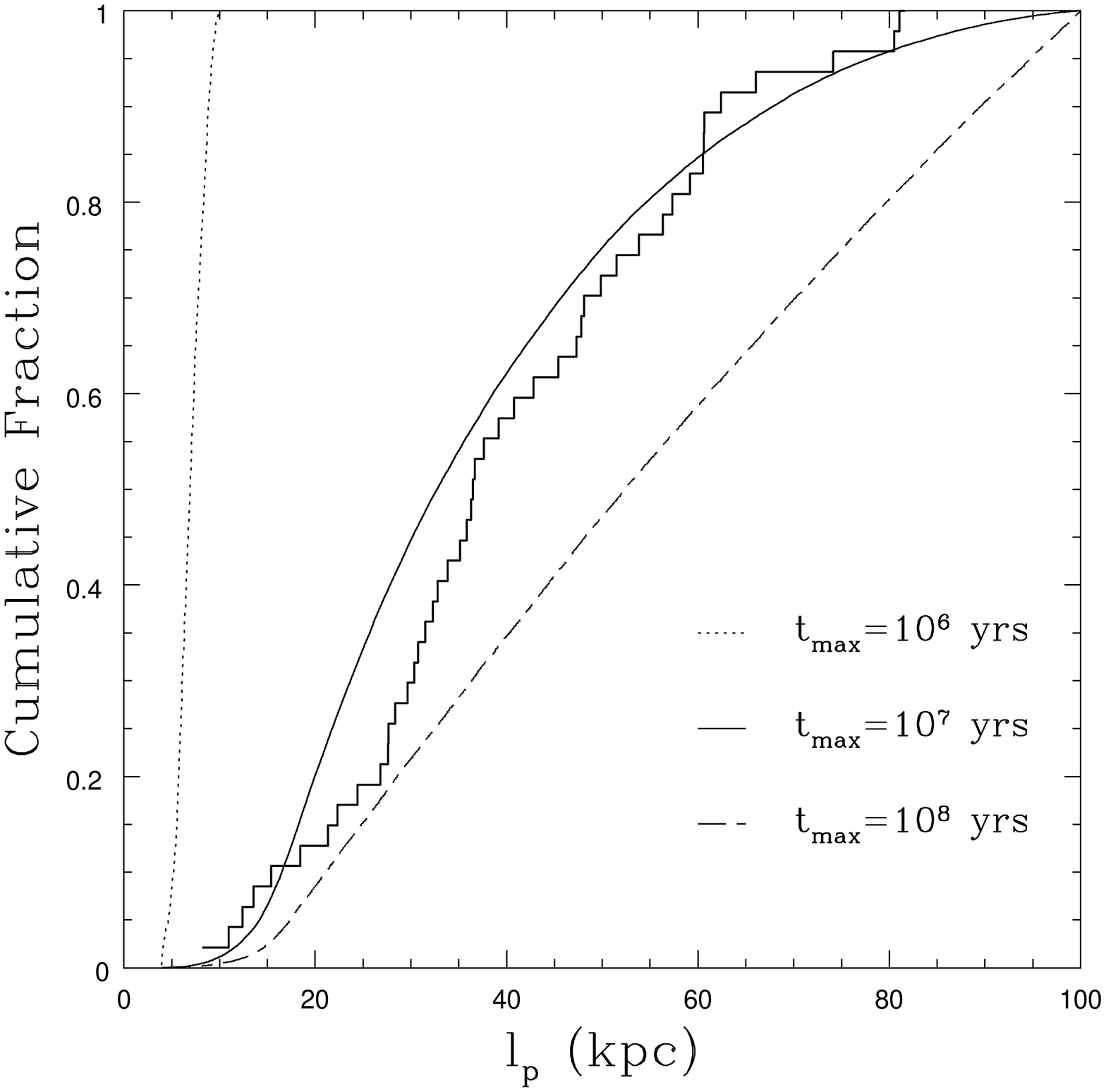}
\caption{\label{fig:cdf_wide}The cumulative distribution of projected lobe lengths in our sample (solid histogram) and representative models with \Tmax$= 10^6$ (dotted line), $10^7$ (solid line), and $10^8$ (dashed line) years. The remaining parameters are our default values, \X{1S} (see Table~\ref{table:LH}). These CDF shapes correspond to equally unique PDFs; thus, we are able to easily determine the most likely \Tmax\ for our observed sample. Of all the variables in our analysis, \Tmax\ has the largest effect on the length distribution. The best fit \Tmax\ $=1.5\times10^7$ years.}
\end{figure}

In order to securely define a sample of FRII sources, we must know both the lobe length and the hotspot to central galaxy separation. We assumed that the centroid of the bright, compact emission in each lobe was the position of the hotspot (as in \citet{Blundell99}). This definition is practical as it traces the position of the brightest emission in each lobe and provides half of the needed information to make FR class distinctions of \citet{Fanaroff74}. Determination of the lobe length is inherently more subjective. We define the lobe length as the maximum separation between the central galaxy and the contour corresponding to $5$ times the average RMS of the FIRST survey ($5*0.15$ mJy bm$^{-1}$). The errors associated with both the position of the brightest emission and the lobe length defined in the above manner are less than one FIRST pixel ($1.8$''). Lobe lengths are measured exclusively using FIRST data as our study is dependent upon accurate astrometry to establish galaxy-lobe associations and to make source length determination as rigorous as possible. The cumulative projected length distribution of the observed sample has a median of 36.5 kpc and is shown in Figure~\ref{fig:cdf_wide}. See Table~\ref{table:Sample} for the projected lobe lengths of the FRII sample.

In our subsequent analysis we only include lobes with a projected length less than $100$ kpc. This cut in lobe length removes five lobes from our FRII sample but suppresses two significant sources of potential error: (1) Longer sources inherently have larger absolute errors in their length measurement and (2) At distances greater than $100$ kpc from the central group galaxy, the potential for associating another group member's radio emission with the central galaxy increases. In addition, we expect at least one convincing fake FRII source (greater than $135^{\degr}$ separation between two random sources with projected length asymmetry less than 15 kpc) at greater distances due to the surface density of FIRST sources. We examine any effect this cut may have on our conclusions in section~\ref{sec:maxage}.

\section{Mock Population Generation}\label{sec:popgen}
To estimate the maximum age of the sources in our sample, we generate mock FRII radio source catalogs that match the selection criteria of our FIRST sample but span a wide range of maximum age. We then compare the model distributions of lobe length with our observed distribution to determine the maximum lifetime \Tmax\ most consistent with the data. There are several considerations one must take into account when creating the mock catalogs: the random projections of our observed jets; the limitations of FIRST (both in angular resolution and sensitivity); and the evolution in luminosity and linear size of radio sources as a function of time.

A simple Monte Carlo scheme is employed to create the mock catalogs. For a given simulation, sources are assigned ages, redshifts, jet powers, and orientations according to the prescriptions in this section.   We detail the age, redshift, and jet power distributions of our catalogs in Section~\ref{sec:global}. Each source is then evolved using the self-similar hydrodynamic KDA model to produce their luminosities and intrinsic lengths. A brief outline of the KDA model and its input parameters follows in Section~\ref{sec:KDA}. Sources are projected onto the sky with a random viewing angle (Section~\ref{sec:geom}). We assume a flat universe with H$_0=71$ \kms, $\Omega_{m}=0.27$, and $\Omega_{\Lambda}=0.73$.

\subsection{Global Population Properties}\label{sec:global}
In order to model the luminosity and length of a mock radio source, the three input parameters are the age, redshift, and jet power. We generate a series of mock catalogs parameterized by the maximum allowed age \Tmax\ of the sources. Each mock source is assigned some time, t$_j$, between its birth time and \Tmax. For a single simulation's sources, t$_j$ is uniformly distributed over the interval [$1$ yr, \Tmax]. In practice it is reasonable to assume some distribution in \Tmax\ as each source will not turn off at exactly the same time. While one could in principle attempt to solve for a spread in \Tmax\ as well, we do not do so because of our small sample size. Introducing a distribution in \Tmax\ to the model would widen the likelihood functions calculated in Section~\ref{sec:res}. Once a source is assigned t$_j$, it is then randomly assigned a redshift from a probability distribution that matches the observed redshift distribution of the parent group sample.

Finally, each source is assigned an initial jet power, $Q$, drawn from a probability distribution of the form:
\begin{equation}
p(Q)dQ \propto Q^{-\alpha}dQ,  \label{Qdist}
\end{equation}
for $Q$ in some range $Q_{min} \le\ Q \le\ Q_{max}$. We adopt two distinct $Q$ distributions, performing a parallel analysis with each. The first is constrained by the local LF of active radio galaxies calculated by \citet{Sadler02}. Jet powers are assigned to our mock sources from a probability distribution such that when evolved with KDA under our default conditions (Table~\ref{table:KDA}), the resultant luminosity function's slope and limits match that of radio emitting AGN \citep[see eq.\ 11,][]{Sadler02}. The Sadler $Q$ distribution (hereafter \Qsad) designates: $Q_{min}=5.0\times10^{33}$ W, $Q_{max}=1.0\times10^{39}$ W, and $\alpha=0.62$. The second $Q$ distribution is taken from \citet{Blundell99}: $Q_{min}=5.0\times10^{37}$ W, $Q_{max}=5.0\times10^{42}$ W, and $\alpha=2.6$ (hereafter \Qblu). While $Q_{max}$ is higher for this distribution, the much steeper slope yields fewer sources with high $Q$. Neither of these $Q$ distributions are ideal for our purposes: The Sadler \Qsad\ is derived relative to an FRI LF, while the Blundell \Qblu\ employed a different cosmology. However we demonstrate below is Section ~\ref{sec:res} and the Appendix that our results are relatively insensitive to the $Q$ distribution. Note that sources within a single simulation are assigned jet powers from only one of these $Q$ distributions.

\subsection{KDA Evolution Model} \label{sec:KDA}
The next ingredient in our mock catalogs is a prescription for radio source  evolution in time. As noted above, three models for FRII source evolution are commonly discussed in the literature: KDA, BRW, and MK. Most of these models only differ in their calculation of the luminosity, whereas the equations governing the length of the radio source as a function of age simply stem from a dimensional argument found in KA. \citet{Barai06} quantitatively compared these models and their ability to reproduce established radio surveys' results, particularly the FRIIs in the Third Cambridge Revised Revised (3CRR), Sixth Cambridge (6CR), and Seventh Cambridge Redshift Surveys (7CRS). These authors created mock catalogs (using a similar scheme to this paper) to emulate these surveys and evaluated each model with 1D and 2D Kolmogorov-Smirnov tests between the real and modeled data across several parameters (luminosity, linear size, and spectral index). They concluded that while all of the models required modification to produce the proper distribution of spectral index, the KDA model produced simulated data that most closely matched the observed surveys, especially in the luminosity-linear size plane most relevant to our investigation. Based on \citet{Barai06}, we use KDA to model the luminosity evolution of FRIIs in our mock catalogs. While we chose the KDA model based on its agreement with existing data, all the models mentioned produce broadly consistent size-luminosity relationships and therefore our results should not significantly depend upon model choice.

KA and KDA also have the virtue of being the simplest self similar models for the length and luminosity of a radio source as a function of age, environment, and jet power. They are an extension of canonical models in which a jet emerges from the AGN region and is soon confined by the uniform pressure of the surrounding cocoon except near the hotspot. The ram pressure of the jet is distributed over the working surface and is balanced by the ambient shocked IGM. KA introduce the dynamics of the cocoon while KDA calculate evolutionary tracks for radio sources in the luminosity-linear size plane. Below we quickly summarize the important assumptions and quantities involved in the calculation of source length.

The KA model assumes the lobe expands into an IGM whose density is parameterized by a simplified King model: $\rho_{x}=\rho_{0}(r/a_0)^{-\beta}$  with $r$ the radial distance from the radio source core, $\rho_0$ a constant density, $a_0$ a scale length, and $\beta$ the radial density index (KA, see their eq.\ 2). Further assuming that the rate of energy injection from the AGN into the cocoon and rest-mass transport along the jet are constant, one can make a dimensional argument that the length of the lobe, $l$, is:
\begin{equation}\label{eq:length}
l=c_1a_0\left(\frac{t}{\tau}\right)^{3/(5-\beta)},
\end{equation}
where
\begin{equation}\label{eq:tau}
\tau\equiv\left(\frac{a_0^5\rho_0}{Q_0}\right)^{(1/3)},
\end{equation}
is a characteristic time scale, $t$ is the age of the source, $Q_0$ is the initial jet power, and $c_1$ is a dimensionless constant of order unity. The calculation of source luminosity is far more complex but still relies upon the basic assumptions outlined above. We use eq.\ 16 of KDA to compute the luminosity, $P_{\nu}$, as a function of time and frequency and refer the reader to that paper for its derivation.

\label{sec:inpars}
Relative to many radio source evolution models, the KDA approach is straightforward. Table~\ref{table:KDA} lists the parameters used in the model, a short description of each, and typical values. Given a choice of these parameters, hereafter collectively referred to as \X, and the age, redshift, and jet power of the source, the KA and KDA models describe the calculations necessary to obtain its luminosity and length.  We first address the parameters that remained constant in our analysis and then discuss those that were varied from simulation to simulation.

The adiabatic index of the gaseous IGM, cocoon, and magnetic field ($\Gamma_x$, $\Gamma_c$, and $\Gamma_B$, respectively) are chosen to represent non-relativistic ($\Gamma_x$) or relativistic ($\Gamma_c$, $\Gamma_B$) equations of state \ie\ $\Gamma_x=5/3$, $\Gamma_c=\Gamma_B=4/3$. There are two other plausible scenarios: both the cocoon and the magnetic field are governed by non-relativistic equations of state or the magnetic energy density is proportional to one of the relativistic particles. Neither scenario would significantly change the length calculation.

Several  jet parameters can float within reasonable ranges. The first of these is the energy distribution of injected particles, which characterizes how energy is transported from the hotspot to the lobe. In the KDA model, radiating particles are injected into the lobe from the hotspot according to some energy number distribution, $N(E) \propto E^{-p}$, and the injection index $p$ is assumed to be constant over the lifetime of the source. In the rest frame of the jet shock, material in the jet is moving with $\Gamma\sim1.67$ (KDA). \citet{Heavens88} determined that this implies an injection index of $p=2.14$. While that approach was not specifically intended for extragalactic shocks, \citet{Alexander87b} and others have shown that typical radio source SEDs imply $2\leq p \leq3$. Therefore, KDA adopts $p=2.14$ as a fiducial value and we do the same.  In section~\ref{sec:p}, we explore the relationship between $p$ and \Tmax, including the effect a more conventional index $p=2.5$ has on the calculated maximum lifetime.

The remaining parameters set the intrinsic properties of the lobe and the surrounding IGM. The axial ratio of the lobe $R_T$ is the ratio of the length of the lobe to its width and is an important variable in the lobe's evolution both in power and as a linear source. The higher the axial ratio of the lobe, the higher the pressure at the hotspot ($p_h$). This higher $p_h$ will produce a greater hotspot advance speed, yielding sources of a given length in a shorter period of time. The distribution of axial ratio in radio sources is still not well characterized by an unbiased survey of FRII sources. Work by \citet{Leahy84} and \citet{Leahy89} at least indicate axial ratios $1.3\le\ \Rt \leq6$. More recent samples confirm this range as nominal for FRII sources \citep[\eg][]{Kharb07}. Ideally, one would assign axial ratios to each model source from a known distribution. However, this distribution is poorly constrained by current observations and we therefore assign all sources a single axial ratio per simulation. KDA adopt a nominal value, \Rt$=2.0$, corresponding to $\theta=31.1^{\circ}$ and we do the same. 

A dense IGM increases the confinement pressure on the lobe and reduces the hotspot advance speed. We investigate how the density profile of the IGM, another input parameter of the KDA model, affects lobe length and lifetime measurements in Section~\ref{sec:den} and determine the relationship between axial ratio and \Tmax\ in Section~\ref{sec:Rt}. 

\begin{deluxetable}{ccl}
\tablecolumns{3}
\tabletypesize{\scriptsize}
\tablecaption{Description and default values for KDA input parameters\label{table:KDA}}
\tablehead{
\colhead{Parameter} &
\colhead{Default Value} &
\colhead{Description} \\
}

\startdata
$\Gamma_x$  &   $5/3$   &   Adiabatic index of the IGM \\
$\Gamma_c$  &   $4/3$   &   Adiabatic index of the cocoon \\
$\Gamma_B$  &   $4/3$   &   Adiabatic index of the magnetic field energy density \\
$p$     &   $2.14$  &   Injection index \\
\Rt     &   $2.0$   &   Axial Ratio \\
$\rho_0$    &   $7.2\times 10^{-22}$ \kgm   &   Constant central density \\
$a_0$       &   $2.0$ kpc   &   Scalelength \\
$\beta$     &   $1.9$   &   Radial density power law index \\
\enddata
\end{deluxetable}

\subsection{Geometric Considerations}\label{sec:geom}

\begin{figure*}[tbp]
\figurenum{3} 
\plotone{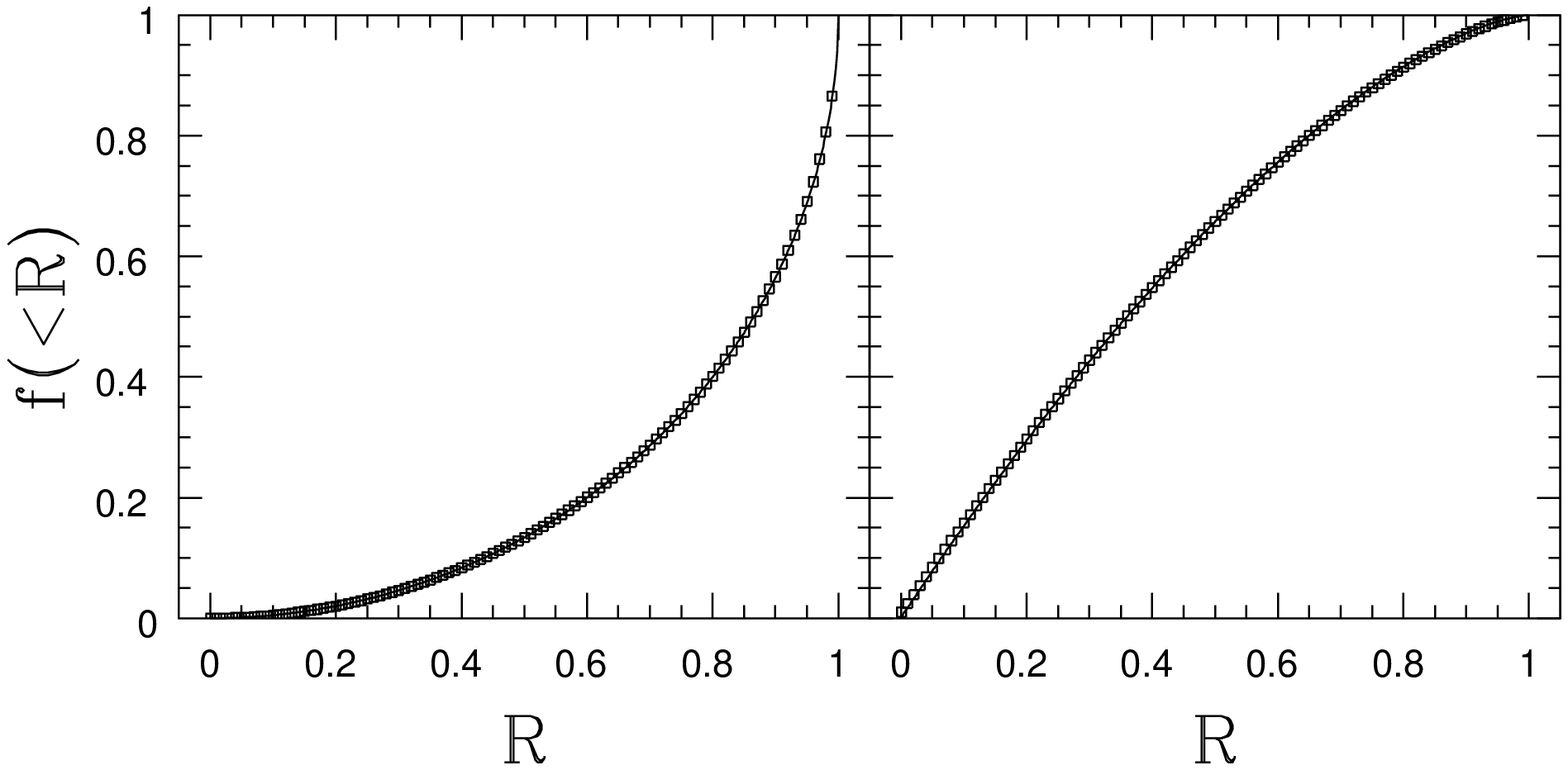}
\caption{\label{fig:proj_length} The cumulative fraction of sources less than a projected length $R$ on the sky. The boxes are the result of random projections of single (left) and uniformly (right) aged sources. The maximum length is normalized to one in both plots. For a population of sources with a single age (left), $50$\% of the sources are projected at $\sim85\%$ or higher of their intrinsic length. Once the population is uniformly distributed in age, the median projected distance is now $36$\% of the intrinsic length of the oldest sources. The boxes are the result of random projections of $10^4$ (left) and $10^5$ (right) sources. The solid lines are given by eq.~\ref{eq:frac_area} (left) and eq.~\ref{eq:multiage} (right).}
\end{figure*}

As our actual measurement is of a distribution of \emph{projected} source length, we finally assign each mock source a random viewing angle. This is more accurate than assuming an average viewing angle for all sources. In the absence of any luminosity evolution, the expected source size distribution for a single age population, as well as for a range of ages up to some \Tmax, can be calculated analytically. For a single age population, where $r$ is the intrinsic source length and $R$ is the projected source length, the fraction of observed sources with length less than R is:
\begin{equation}\label{eq:frac_area}
f(<R)=1-\sqrt{1-\left(\frac{R}{r}\right)^2}.
\end{equation}
This function is shown in the left panel of Figure~\ref{fig:proj_length}, along with the projected length distribution of $10^4$ random sources of a single age. Each source is assigned a random $\theta$,$\phi$ and projected on the sky, yielding the length distribution shown by the data points. Our random distribution is in excellent agreement with the analytic result.

We now extend this test to a set of sources uniformly distributed in age up to some maximum age \Tmax\ associated with a maximum intrinsic length $r_{max}$. The fraction of sources less than some projected length $R$ is:
\begin{equation}\label{eq:multiage}
f(<R)=\int_0^R dr + \int_R^{r_{max}}1-\sqrt{1-\left(\frac{R}{r_{max}}\right)^2} dr.
\end{equation}
We naturally split the equation between contributions from sources with intrinsic length $r<R$  and $r\ge R$. For all $r<R$, $f(<R)$ picks up the number of sources with intrinsic length $r$. For other values of $r$, we sum over the fraction of sources less than $R$ for each intrinsic length. The right side of Figure~\ref{fig:proj_length} shows $f(<R)$ for a $10^5$ source sample (data points) overlaid with eq.~\ref{eq:multiage}. Notice how the projected length distribution is shifted towards smaller source lengths, with the median source length now at $36\%$ of $r_{max}$. In practice, this analytic distribution is not a good match to the data because of the implicit assumption of no luminosity evolution. Because sources fade as they age, the observed distribution should be skewed towards shorter sources.

\subsection{Catalog Generation}
For a given choice of parameters \X\ (see Table~\ref{table:KDA}) and maximum age \Tmax, we generate a mock catalog with $10^6$ radio sources. Each source is assigned a jet power, redshift, and age according to the distributions detailed above. The intrinsic length and luminosity $P_{\nu}$ of each source is obtained using the KDA model. These are randomly projected on the sky as described in section~\ref{sec:geom}, yielding projected length $l_p$. The mock catalogs are stripped of undetectable sources via length and luminosity cuts to match the observational limits of the FIRST survey. If $l_p$ is less than our $10$'' minimum angular size for detected FRIIs, it is removed from the final catalog. Similarly, if $P_{\nu}$ corresponds to a flux below the 1 mJy sensitivity of FIRST, it is discarded. We examined a similar catalog generation process using a surface brightness cut and noted no statistical difference in the resultant mock catalogs when compared to those created using the FIRST flux limit.  This demonstrates that our analysis is not biased against large, diffuse, presumably old lobes that might have been resolved out by FIRST's exquisite angular resolution. The output from this process is a cleaned catalog of sources whose properties are well understood and meet our selection criteria.

\section{Results: Lifetime of FRIIs}\label{sec:res}
If the physical properties of a set of radio lobes are kept constant, their maximum age sets the length distribution. In the next subsections we compare the projected length distribution of the models discussed in Section ~\ref{sec:popgen} with the observed distribution and quantify the goodness of fit via a maximum likelihood estimation (MLE). We first describe our maximum likelihood approach in Section~\ref{sec:maxage} and use this framework to estimate \Tmax. We then investigate degeneracies between \Tmax\ and other parameters of the KDA model in Section~\ref{sec:pars}.

\subsection{Method of Maximum Source Age Determination}\label{sec:maxage}
We use the maximum likelihood approach to determine which  models best reproduce the length distribution observed. In general a model is made distinct by a unique choice of \Tmax\ and model parameters \X{i}\ (Section ~\ref{sec:inpars} and Table~\ref{table:KDA}). Every choice of \X{i}\ and \Tmax\ produces a distinct distribution of projected lobe length $l_p$.

We first create probability density functions (PDFs) of the model distributions we wish to compare. For each mock catalog, we divide the one dimensional space of $l_p$ into equally wide bins. The probability, $\mathcal{P}_k$, assigned to the $k$th bin is then just $n_k/\sum_{j}n_j$ where $n_k$ is the number of sources within the $k$th bin and $\sum_{j}n_j$ is the total number of sources in all the bins. The likelihood, $\mathcal{L}$, of a model is then:
\begin{equation}
\mathcal{L}=\prod_{j}\mathcal{P}_j,
\end{equation}
where $\mathcal{P}_j$ is the probability of finding the $j$th source in the model. As the PDFs are relatively smooth functions, we use simple linear interpolation to obtain $\mathcal{P}_j$ from $\mathcal{P}_k $.

In order to reasonably limit the theoretical error of our model with a feasible number of sources, we only evaluated the probability of observed sources with $5.3 \le\ l \le\ 101.3$ kpc. Using these length restrictions, we are assured good statistics in our longest length bins and avoid potential sources of error such as spurious associations (Section~\ref{sec:meas}) and sources that fall below the angular resolution of FIRST at the highest redshift of the parent sample ($z=0.1$). The maximum length of 101.3 kpc was simply chosen to produce 33 equally spread bins of 3 kpc each.  As we show below, these cuts in length do not limit our sensitivity to maximum age distinctions because the observed length distribution is such an important constraint (see Figure~\ref{fig:cdf_wide}). We create $10^6$ sources per simulation to ensure uniform sampling of the chosen distributions of jet power, age, and redshift.

The prescription for determining the lifetime of FRII radio sources in now straightforward. For a fixed choice of \X{i}, we produce models with a range of \Tmax\ and determine that which maximizes the likelihood. Specifically, models with $2\times10^4$ sources were created for a range of \Tmax\ between $10^6$ and $10^8$ yrs in steps of $10^6$ yrs. Once the approximate peak in the likelihood function is found, we produce models with $10^6$ sources, making smaller steps in \Tmax\ around the peak. The lifetime of FRII sources for a given \X{i} is then the \Tmax\ of the model of maximum likelihood. One $\sigma$ errors, corresponding to $\Delta\ln \mathcal{L} =0.5$, are calculated in each direction. Table~\ref{table:LH} lists the lifetime of FRII sources for a variety of \X{i}.

\begin{deluxetable*}{lcccccccc}
\tablecolumns{9}
\tablewidth{6.0truein}
\tabletypesize{\scriptsize}
\tablecaption{Maximum Lifetime Estimates}
 \tablehead{
\colhead{ID} &
\colhead{$\rho_0$} &
\colhead{$a_0$} &
\colhead{$\beta$} &
\colhead{\Rt} &
\colhead{$p$} &
\multicolumn{2}{c}{---------\Tmax ($10^7$ yr)---------} &
\colhead{Notes} \\
\colhead{} &
\colhead{(\kgm)} &
\colhead{(kpc)} &
\colhead{} &
\colhead{} &
\colhead{} &
\colhead{\Qsad} &
\colhead{\Qblu} &
\colhead{} \\
\colhead{(1)} &
\colhead{(2)} &
\colhead{(3)} &
\colhead{(4)} &
\colhead{(5) } &
\colhead{(6)} &
\colhead{(7)} &
\colhead{(8)} &
\colhead{(9)} \\
}

\startdata

\X{1S}, \X{1B} & $7.2\times10^{-22}$ &  2.0 &  1.9 &  2.0 &  2.14& $1.22\pm^{0.28}_{0.14}$ & $1.75\pm^{0.21}_{0.07}$& default (KDA)\\
\X{2S}, \X{2B} & $1.67\times10^{-23}$ &  10.0 &  1.5 &  2.0 &  2.14& $1.15\pm^{0.22}_{0.15}$& $1.59\pm^{0.19}_{0.09}$& BRW density profile\\
\X{3S}, \X{3B} & $7.19\times10^{-26}$ &  391.0 &  1.23 &  2.0 &  2.14& $0.88\pm^{0.20}_{0.06}$& $1.31\pm^{0.14}_{0.10}$& \citet{Jetha07} density profile\\
\X{4S}, \X{4B} & $7.2\times10^{-22}$ &  2.0 &  1.9 &  1.0 &  2.14& $2.44\pm^{0.51}_{0.27}$& $3.55\pm^{0.37}_{0.22}$& \Rt=1.0\\
\X{5S}, \X{5B} & $7.2\times10^{-22}$ &  2.0 &  1.9 &  4.0 &  2.14& $0.72\pm^{0.14}_{0.09}$& $1.04\pm^{0.09}_{0.04}$& \Rt=4.0\\
\X{6S}, \X{6B} & $7.2\times10^{-22}$ &  2.0 &  1.9 &  6.0 &  2.14& $0.53\pm^{0.12}_{0.06}$& $0.76\pm^{0.09}_{0.04}$& \Rt=6.0\\
\X{7S}, \X{7B} & $7.2\times10^{-22}$ &  2.0 &  1.9 &  2.0 &  2.5& $1.26\pm^{0.26}_{0.14}$& $1.79\pm^{0.13}_{0.09}$& p=2.5\\
\X{8S}, \X{8B} & $7.2\times10^{-22}$ &  2.0 &  1.9 &  2.0 &  2.9& $1.26\pm^{0.32}_{0.15}$& $1.79\pm^{0.19}_{0.12}$& p=2.9\\

\enddata

\tablecomments{\label{table:LH} FRII lifetimes for a variety of jet, lobe, and IGM conditions. (1): Identifier of the parameter choice in the text. \X{iS}\ is the parameter set constrained by the Salder LF; \X{iB}\ is constrained by the BRW jet power distribution. (2): IGM density at the core radius in units of \kgm. (3): Scale radius in kpc of the King profile used to model the IGM. (4): Index of density profile: $\rho(r)=\rho_0(r/a_0)^{-\beta}$. (5): Axial ratio of lobes in simulation. (6): Injection index of energy number distribution of particles in the jet: $N(E)\propto E^{-p}$. (7): Most likely \Tmax\ for \X{iS}\ simulation in $10^7$ years. (8): Most likely \Tmax\ for \X{iB}\ simulation in $10^7$ years. (9): Brief description of how the parameter set differs from the default set.}
\end{deluxetable*}

\begin{figure}[bp]
\epsscale{1.10}
\figurenum{4} 
\plotone{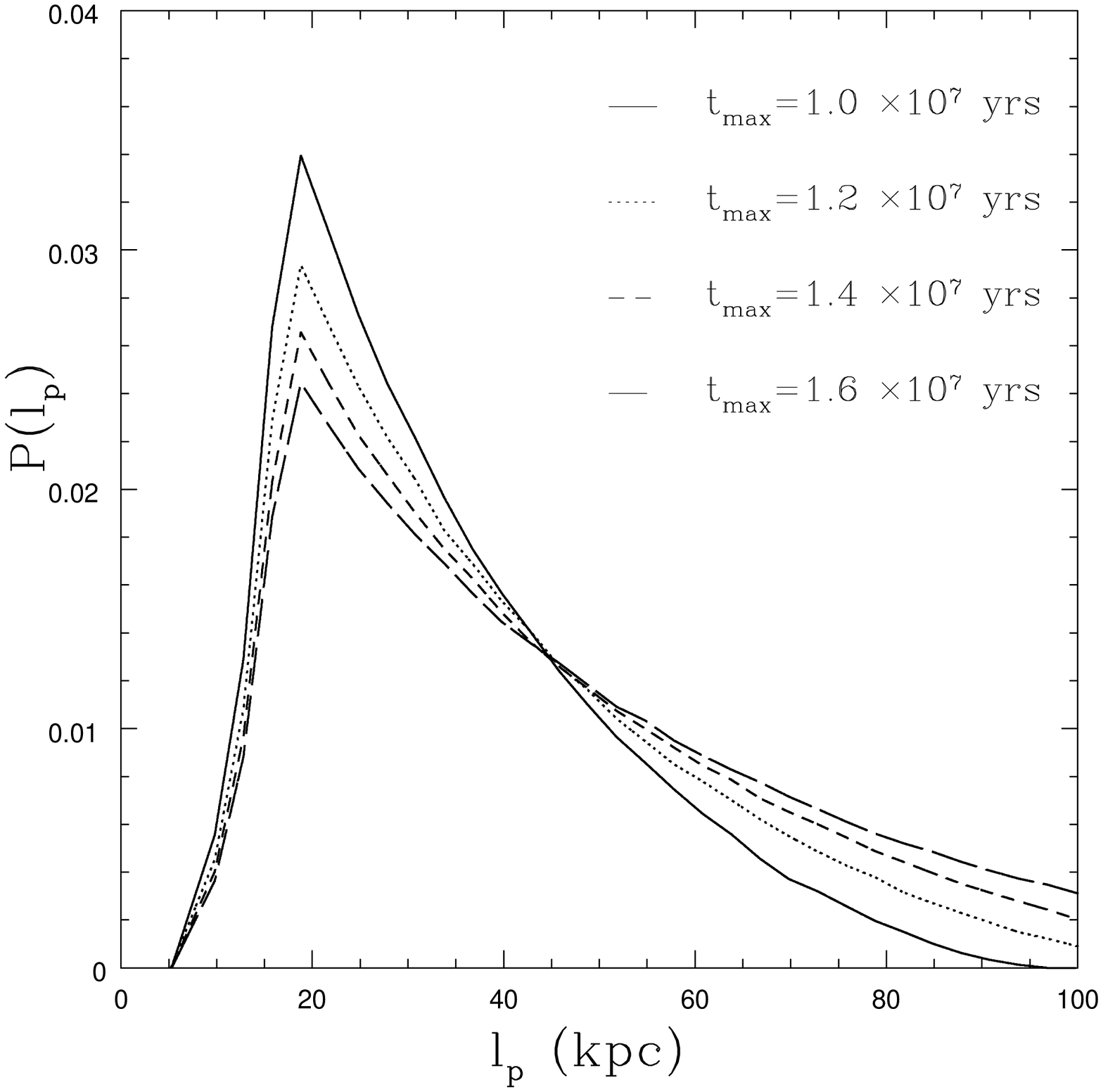}
\caption{\label{fig:pdf_vary}PDFs for several values of \Tmax: $1.0\times10^7$ (solid line), $1.2\times10^7$ (dotted line), $1.4\times10^7$ (short dashed line), $1.6\times10^7$ (long dashed line). Default values were used for all input parameters to the KDA model (\X{1S}, see Table~\ref{table:LH}).}
\end{figure}

Figure~\ref{fig:cdf_wide} shows the cumulative length distribution of our sample, along with mock catalogs generated with our default KDA input parameters (\X{1s}) for \Tmax$= 10^6, 10^7,$ and $10^8$ years. It is important to note how dramatically different the model distributions are for different \Tmax. The fact that these distributions are so distinct allow us to determine \Tmax\ with good precision and makes this technique so effective. For \X{1S} and \Tmax$=10^6$ yrs, the median projected lobe  length is $\sim 7$ kpc, yielding no sources $\ge 11$ kpc. Conversely, a lifetime of $10^8$ yrs produces of median source length of $\sim 53$ kpc. Shorter \Tmax\ skews the distribution towards young sources that do not have enough time to produce the longer observed lengths, whereas most sources are older than $10^7$ years in the \Tmax$= 10^8$ yrs case and therefore the expected observed distribution is much more heavily weighted towards longer lobes. Figure~\ref{fig:pdf_vary} shows the PDFs of several \Tmax\ for \X{1S}. Even $\Delta$\Tmax$\sim 2\times10^6$ years produces appreciable changes in the PDF and thus the likelihood, $\mathcal{L}$. The \Tmax\ best representing the data, the maximum of the $\mathcal{L}$(\Tmax) function, is obvious given a choice of \X.


\begin{figure*}[tbp]
\figurenum{5} 
\plotone{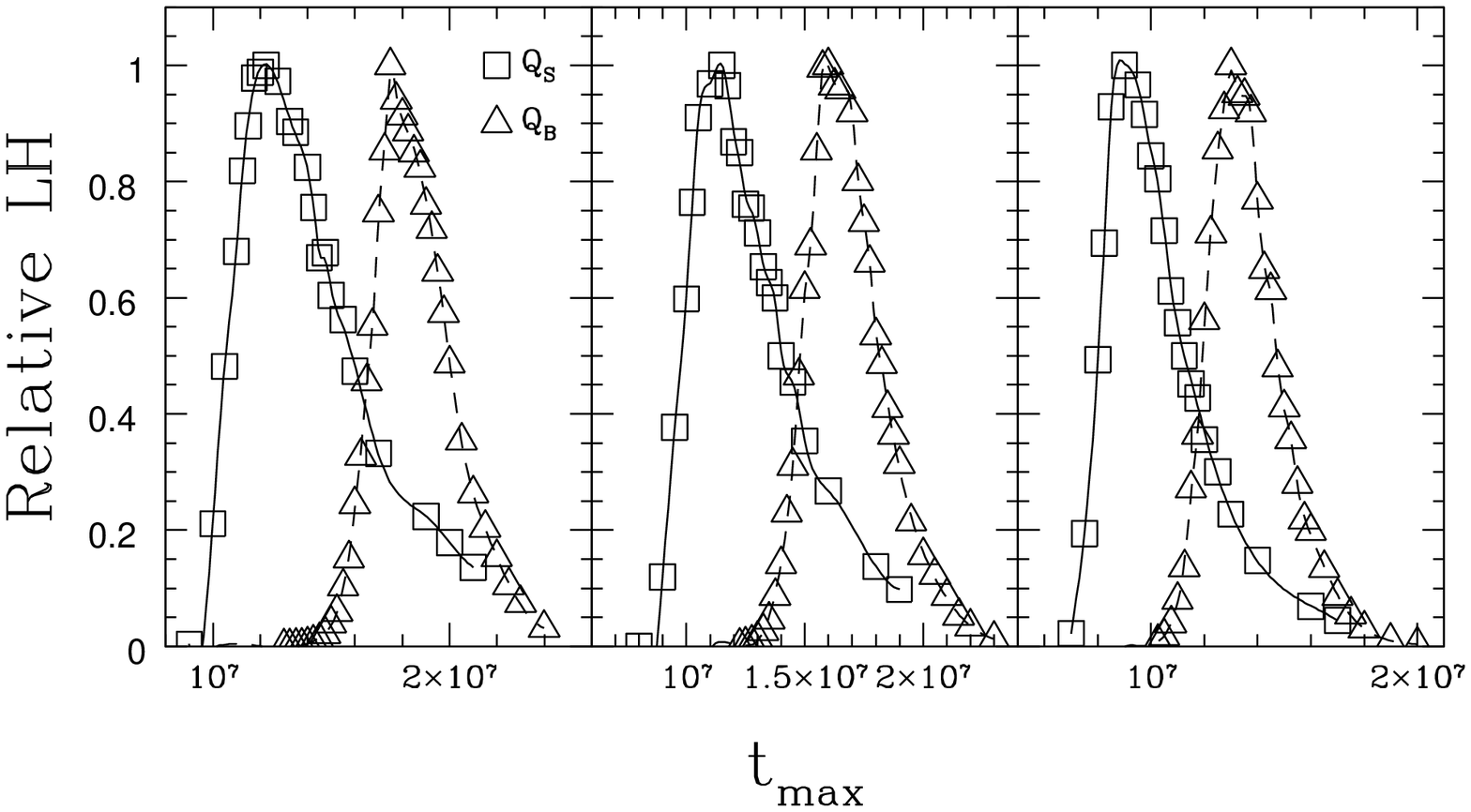}
\caption{\label{fig:den_LH}The relative likelihood as a function of \Tmax\ for our default density profile (left), the BRW IGM density profile (middle), and the IGM density profile of \citet{Jetha07} (right). The peak in the relative likelihood curve represents the most likely \Tmax\ given our observed sample. The input parameters of these simulations were \X{1}, \X{2}, and \X{3}\ (left, middle, and right). The squares (triangles) are the relative likelihoods for simulations run with the \Qsad\ (\Qblu\ ) distribution. The solid and dashed lines are spline fits used to calculate the error of the maximum likelihood estimate.}
\end{figure*}


The results for our fiducial parameter sets, \X{1S,1B}\ are shown in Table~\ref{table:LH}. For \X{1S} the most probable \Tmax$= 1.22(\pm^{0.28}_{0.14})\times10^7$ yrs, while for the Blundell distribution \X{1B} the most likely \Tmax$=1.75(\pm^{0.21}_{0.07})\times10^{7}$ yrs. A plot of the relative $\mathcal{L}$ as a function of \Tmax\ for \X{1S}\ and \X{1B}\ is shown in the left panel of Figure~\ref{fig:den_LH}. These two maximum age estimates are comparable because the median jet powers of the two $Q$ distributions are similar in spite of their different ranges and slopes. The Blundell $Q$ distribution has a slightly lower median jet power, hence the longer estimated lifetime of sources.

\subsection{Parameter Dependence}\label{sec:pars}
We have determined the FRII lifetime for a fiducial choice of model parameters (section~\ref{sec:maxage}). An immediate concern is how degenerate the FRII lifetime is with various properties of the jet, lobe, and ambient IGM. We produce a series of models \X{i}, representing the extremes of the observed parameter space discussed in Section~\ref{sec:inpars}. We estimate the lifetime of FRII sources with these parameter choices to place constraints on the systematic uncertainties due to the unknown values and distributions of these parameters. As these models should bracket the true range of jet, lobe,  and environment properties, we expect the corresponding range of \Tmax\ will be a conservative estimate of its uncertainties. See section~\ref{sec:sum_par} for a brief explanation of the choice of singular values over distributions of \Rt\ and density profile in a given simulation. An analytic investigation of parameter dependence is presented in the Appendix.

\subsubsection{Density Profile}\label{sec:den}
The density of the IGM surrounding  a lobe could have a significant impact upon its length. This introduces a possible degeneracy between a chosen density profile of the IGM and the lifetime we measure for our FRII sources.  As mentioned in Section~\ref{sec:KDA}, KDA approximates the density profile of the IGM with a modified King model ($\rho=\rho_{0}(r/a_0)^{-\beta}$). Here, we examine three density profiles and their affect on the maximum lifetime of FRII sources. Two profiles (those of \X{1} and \X{2}) are the parameters used in the KDA and BRW models for radio source evolution. We also implement an empirically derived density profile of the IGM in local groups \citep[][\X{3} in Table~\ref{table:LH}]{Jetha07}.

Our first density profile is associated with the default parameters of the KDA model. The average density $\rho_0$ at the scale radius $a_0$ is $7.2\times10^{-22}$ \kgm, or less than 1 hydrogen atom per cubic meter. \citet{Canizares87} argue that these values are typical out to a radius of $\sim 100$ kpc from the source core. The value of $\beta=1.9$ is larger than that of the other density profiles discussed here but is still within the expected range. \citet{Falle91} showed that for $\beta>2$ the jet will not form shocks and thus no FRII sources will be present. KDA use this choice of beta as it agrees with early X-ray observations of galaxies at low redshift \citep{Cotter96}. We adopt $\rho_0=7.2\times10^{-22}$ \kgcm, a$_0=2.0$ kpc, and $\beta=1.9$ as our fiducial density profile. As noted previously, we calculate a lifetime of $1.22(\pm^{0.28}_{0.14})\times10^7$ yrs for \X{1S} and $1.75(\pm^{0.21}_{0.07})\times10^{7}$ yrs for \X{1B} (Figure~\ref{fig:den_LH}, left panel).

In addition to KDA, the BRW model has also been used with success to explain the evolution of FRII sources with time. The BRW model adopts values of $\rho_0=1.67\times10^{-23}$ \kgm, a$_0=10$ kpc, and $\beta=1.5$. \citet{Garrington91} use the depolarization of polarized synchrotron emission from the lobes of radio galaxies to suggest that radio galaxies preferentially reside in poor group environments. BRW ascertain a density profile from ROSAT observations of such groups \citep{Mulchaey98, Willott98}. Again, the average density at the scale length is less than one atom of hydrogen per cubic meter. It stands to reason that this averaged, measured density profile is a good representation of the IGM in poor groups. We therefore perform our MLE using the BRW density profile (\X{2}). Our likelihood as a function of \Tmax\ is shown in the middle panel of Figure~\ref{fig:den_LH}. The most likely \Tmax\ is $1.15(\pm^{0.22}_{0.15})\times10^{7}$ yrs (\Qsad, \X{2S}) and $1.59(\pm^{0.19}_{0.09})\times10^{7}$ yrs (\Qblu, \X{2B}). These values are in good agreement with those obtained with the KDA density profile. To good approximation, the KDA model depends upon the quantity $\rho_{0}a_{0}^{\beta}$ rather than independent variations in each term.  The BRW density profile product is a factor of five less than that of KDA yet only changes \Tmax\ by a factor of $\sim0.8$.

With the advent of high resolution X-ray observations with {\it Chandra}, gas densities in galaxies and groups are being precisely measured as a function of distance from the galaxy core. While one would like to know the distribution of gas density profiles seen in groups and clusters, the best available data are density measurements of group galaxies by \citet{Jetha07}. Using {\it Chandra}, they looked at 15 nearby groups of galaxies and computed a best fit power law to the radial gas temperature and density profiles of these groups. We compute the maximum lifetime for this density profile (parameter set \X{3} in Table~\ref{table:LH}) and find \Tmax$= 8.8(\pm^{2.0}_{0.6})\times10^6$ yrs for \Qsad\ and $1.31(\pm^{0.14}_{0.10})\times10^7$ yrs for \Qblu\ (see also Figure~\ref{fig:den_LH}, right panel). The Jetha density profile product is 24 times smaller than our default and alters \Tmax\ by a factor of $\sim0.7$. The similar maximum ages calculated with these three density profiles strongly suggests our results are a weak function of IGM density profile. We further explore this behavior in the Appendix.

\subsubsection{Axial Ratio}\label{sec:Rt}
The axial ratio characterizes the shape of the lobe. More collimated lobes reach longer distances in a shorter amount of time as the area of their hotspot is smaller, allowing the jet's ram pressure to build up faster. We note that $1.0\le \Rt\ \le6.0$ have been found for FRII sources \citep{Leahy84, Leahy89}, although  \citet{Machalski04} gathered a sample of radio sources from the literature and found axial ratios as high as 7.5. However, this sample was biased towards the most powerful radio sources known and may not be indicative of typical FRIIs. \citet{Kaiser99} found that \Rt\ is dependent upon $\beta$ while BRW and others postulate that axial ratio is a function of jet power. Either scenario would help explain some of the high axial ratios found in \cite{Machalski04}. As the vast majority of FRII sources appear to have axial ratios between 1.0 and 6.0, we focus our attention on this region of parameter space.

\begin{figure}[tbp]
\figurenum{6} 
\epsscale{1.1}
\plotone{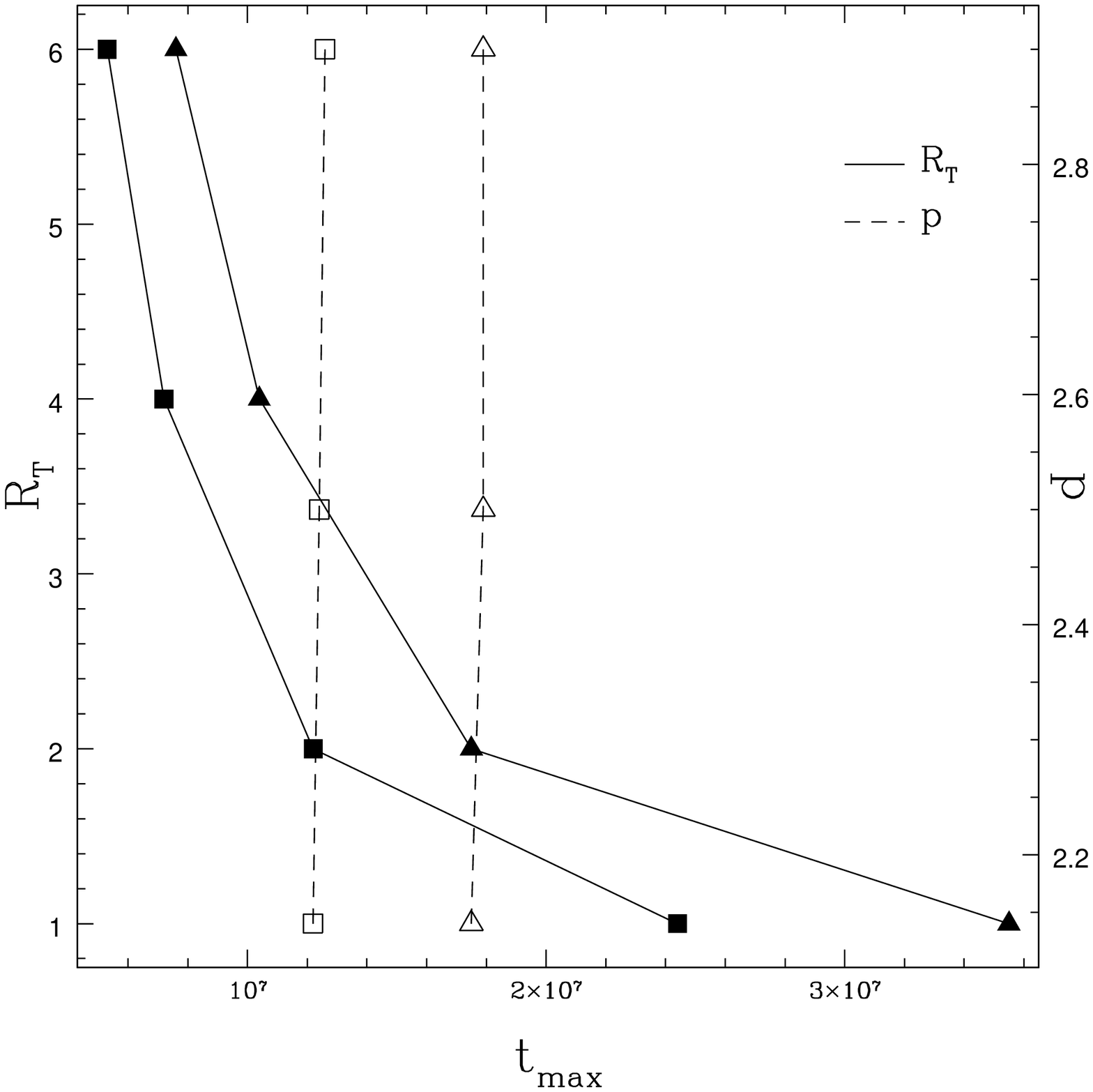}
\caption{\label{fig:Rt_p}Maximum age \Tmax\ as a function of \Rt\ (solid symbols and lines, left ordinate) and $p$ (open symbols and dashed lines, right ordinate). Except for those parameters that are explicitly varied to test their degeneracy with \Tmax, default values of the KDA model input parameters are used. Both the \Qsad\ (squares) and \Qblu\ (triangles) jet power distributions are shown. Note that \Tmax\ shows no dependence on $p$. \Rt\ has a strong effect on \Tmax\, but is still within a factor of $\sim2$ of the default value even at the extremes of the tested range.}
\end{figure}

In addition to our default value of \Rt$=2.0$, we also investigate \Rt$=1,4,$ and $6$. Our \LH{}(\Tmax) for all eight of these parameter choices is shown in Figure~\ref{fig:Rt_p}. When \Rt\ decreases, the pressure at the hotspot increases as the lobe encounters a higher resistance from the IGM due to the increased surface area of the cocoon (KDA). Consequently, the hotspot advance speed decreases and we therefore expect the FRII lifetime to be inversely related to \Rt. If the lobes are less collimated, sources must be older to reach a given linear size. We refer the reader to Table~\ref{table:LH} for our computed values of the FRII lifetime for \Rt\ $=1.0, 2.0, 4.0$, and $6.0$. 

\subsubsection{Injection Index}\label{sec:p}

The injection index $p$ (Section~\ref{sec:inpars}) is used solely in the calculation of the source luminosity. We refer the reader to KDA for an explanation of the relationship between luminosity and $p$. Here, we test the stability of our results against changes in the power law describing the injection of energetic particles into the jet. We have noted that substantial theoretical and observational evidence exists for $2.0\le$p$\le3.0$  (Section~\ref{sec:inpars}). To sample this range, we examine $p=2.5$ and $2.9$ in addition to our fiducial value of $p=2.14$. Recall that the lifetime for FRII sources under our default conditions ($p=2.14$) was $1.22(\pm^{0.28}_{0.14})\times10^7$ yrs (\X{1S}) and $1.75(\pm^{0.21}_{0.07})\times10^7$ yrs (\X{1B}). Upon changing the injection index to 2.5 (\X{7S}), the most likely \Tmax\ is $1.24(\pm^{0.26}_{0.14})\times10^7$ yrs; $p=2.9$ (\X{8S}) yields a lifetime of $1.26(\pm^{0.32}_{0.15})\times10^7$ yrs (Figure~\ref{fig:Rt_p}). Using the Blundell jet power distribution, we measure lifetimes of $1.79(\pm^{0.13}_{0.09})\times10^7$ yrs (\X{7B}) and $1.79(\pm^{0.19}_{0.12})\times10^7$ yrs (\X{8B}). While variation of $p$ should not greatly influence radio luminosity at a given frequency, we have confirmed that changing the value of $p$ within the theoretically accepted range produces no significant change in the lifetime estimate.

We assert that our results are pseudo independent of source luminosity due to FIRST's sensitivity at this observing frequency and the expected luminosity range of FRIIs. Unless other models compute source luminosity in drastically different ways, we expect our results to be robust to changes in radio source evolution model. In particular, as the models of BRW and MK are based upon the KA model, use of either of these models should produce no appreciable changes to these results.

\subsubsection{Summary}\label{sec:sum_par}
We have measured the range of \Tmax\ consistent with our observed sample given the uncertainties in the parameters that characterize the parent population. Estimates of \Tmax\ vary from $5.3\times10^6$ to $3.55\times10^7$ yrs. Within the accepted range of input parameters described above, the most uncertain variable is \Rt. Extreme lobe axial ratios, compared to our fiducial value, produce a factor of 2 change in \Tmax. 

While one could allow for a distribution in density profile and axial ratio as part of this analysis, we judge that our present sample size is too small to provide meaningful constraints. Instead, we have examined a wide and reasonable range of both density profile and axial ratio. If distributions (as yet unquantified by observations) in these parameters were used instead, the resulting \Tmax\ would be a weighted mean of the results already obtained here. When generally discussing FRII lifetime in subsequent sections, we conservatively choose \Tmax$= 1.5(\pm0.5)\times10^7$ yrs. Other choices of \X{i} do not alter the lifetime by more than $20\%$ compared to those found with \X{1S} and \X{1B}.

\subsection{FRII Duty Cycle}
In addition to their lifetime, the duty cycle associated with FRIIs is an important timescale. Using the \citet{Berlind06a} catalog of groups, we measure the typical time between episodes of FRII radio activity in central group galaxies. Of the $2020$ groups studied, $26$ central galaxies were host to FRII radio sources. We determine the level of incompleteness from our mock catalogs. The default simulation with \Qsad\ (\X{1S}) yielded a lifetime of $1.2 \times 10^7$ years.  Of the $10^6$ sources in \X{1S}, $6\times10^5$ would have been detected by FIRST and identified as FRIIs. We therefore estimate that our observed sample of FRII sources is $60\%$ complete for sources that follow the quoted model parameters (\eg\ those with initial jet powers above $Q_{min}$). Sources may not be detected for two reasons: (1) Too young and therefore smaller than the angular resolution of FIRST or our required angular size to be considered an FRII and (2) Luminosity corresponds to a flux below FIRST's sensitivity at the source redshift. We conclude that the incompleteness corrected FRII duty cycle is $2.2\%\ (44/2020)$ in central group galaxies. The corresponding average time between FRII phases in these central galaxies is $5.6 \times 10^8$ years. Due to a different completeness level for \Qblu\ (\X{1B}), $1.7\%$ of central galaxies exhibit FRII activity using this jet power distribution. The time between FRII phases is thus a factor of 2 higher at $1.0 \times 10^8$ years. We note that there are many FRI sources in this group sample as well and consequently the environments of some of these sources may not be amenable to an FRII morphology. This implies the FRII duty cycle is actually higher for those group environments suitable for FRIIs.

\section{Results: Population Statistics and Clustering}
\label{sec:res_clustering}
Our attention was initially brought to the Berlind group sample \citep{Berlind06a} by their subsequent paper on the clustering or bias of groups as a function of various group and central galaxy properties \citep{Berlind06b}. It is interesting to determine if these properties are correlated in any way with radio activity, and whether radio activity is correlated with more clustered systems. Here we quickly outline the results of \citet{Berlind06b} and discuss how radio activity varies with a number of the parameters studied therein.

Group and central galaxy properties tested for correlations with bias in \citet{Berlind06b} were group number richness $N_{grp}$, group velocity dispersion $\sigma_v$, total group color (g-r)$_{tot}$, central galaxy color (g-r)$_{cen}$, central galaxy luminosity M$_{r,cen}$, and the magnitude gap between the central galaxy and the second brightest member of the group $\Delta$M$_r$. Assuming group luminosity and mass were monotonically related and matching the group LF to the halo mass function of \citet{Warren06}, \citet{Berlind06b} divided their sample into four bins of group mass centered at $10^{12.5}$, $10^{13.0}$, $10^{13.5}$, and $10^{14.0}$ h$^{-1}$M$_\odot$. For every galaxy or group parameter, each mass bin was split into  ``high'' and ``low'' subsets, based upon the median parameter value in a given mass bin. The ratio of the high and low bias functions (see eq.\ 3 of \citet{Berlind06b}) was then evaluated for each property across the mass bins. They found that only $N_{grp}$ and $\sigma_v$ were strongly correlated with clustering over their entire mass range and conclude that bluer galaxies are located in \emph{more} strongly clustered groups than redder central galaxies. For the most massive systems, they also found $\Delta M_r$ is a function of bias while they reported no significant correlation between clustering and M$_{r,cen}$. In light of these results, we examine the trends between radio activity and each of these parameters.

Our sample of $2020$ central group galaxies is comprised of the most luminous groups in the Berlind catalog. We determine the radio fraction of central galaxies ($r_{frac}$) with three distinct definitions of what constitutes a radio source: the central galaxy simply contains a core source or is host to extended radio emission (either FRI or FRII) (\rfraca); the radio source has $L_{FIRST} \ge 10^{22.5}$ W Hz$^{-1}$ (\rfracb) as in \citet{Best06a}; and most conservatively, the central galaxy is radio loud with $\log F_{FIRST} / \log F_{SDSS} \ge 1$ \citep[\rfracc, ][]{Ivezic02}. By any of these definitions, a larger \rfrac\ corresponds to a greater likelihood of harboring an active radio source. We broadly expect larger, more clustered systems to have a larger viral radius and possibly more virialized gas than smaller systems. If radio mode feedback is indeed a mechanism to combat the cooling flow problem, one might expect a correlation between \rfrac\ and bias.

\begin{figure*}[tbp]
\figurenum{7} 
\plotone{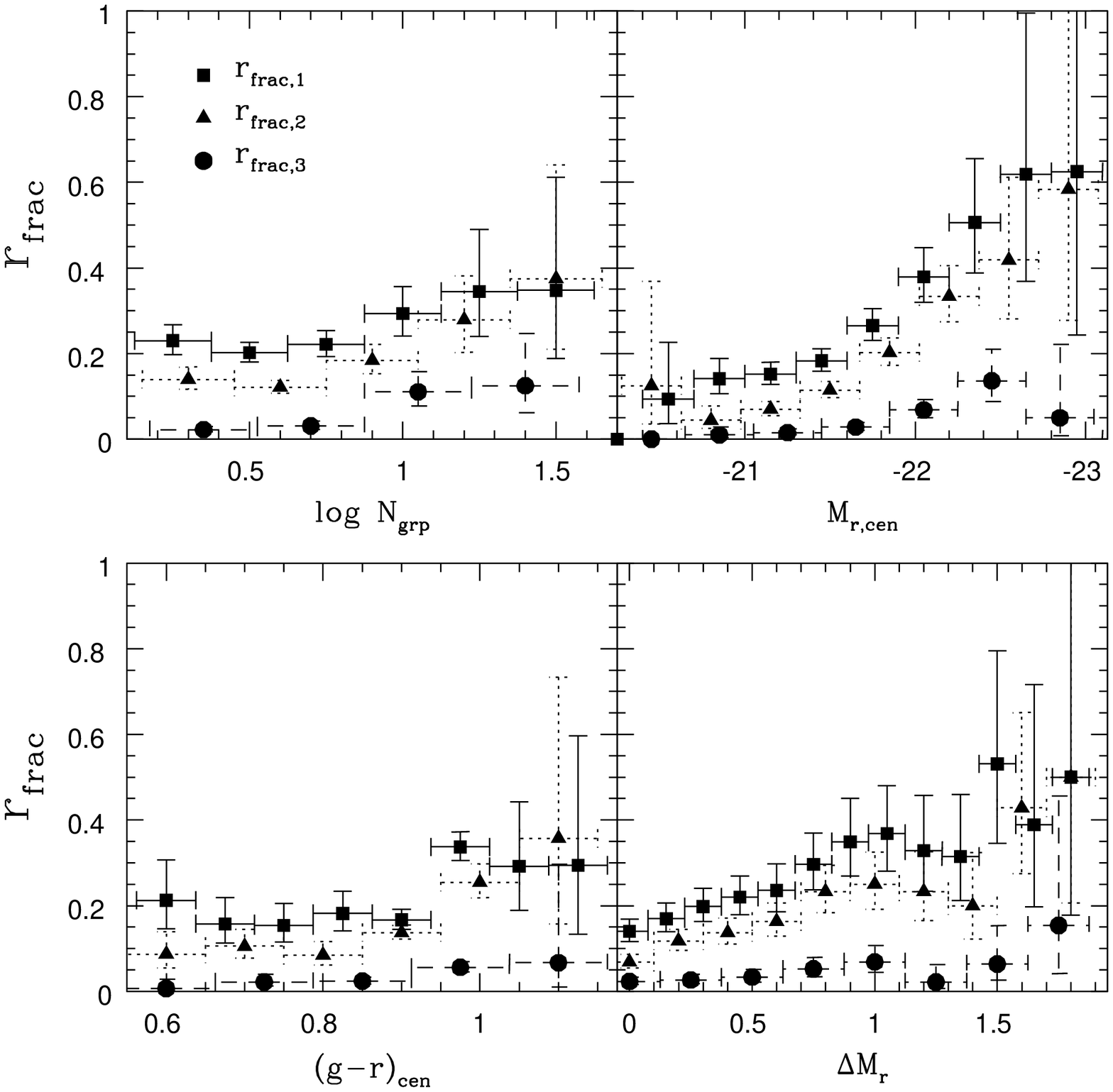}
\caption{\label{fig:clustering} Radio fraction as a function of $N_{grp}$ (upper left), M$_{r,cen}$ (upper right), (g-r)$_{cen}$ (lower left), and $\Delta M_r$ (lower right). The three \rfrac\ criteria in the text are shown: \rfraca\ (squares), \rfracb\ (triangles), and \rfracc\ (circles). Vertical error bars are Poisson; horizontal represent bin width.}
\end{figure*}

We first determine the radio fraction as a function of group richness. Figure~\ref{fig:clustering} (upper left) shows \rfrac\ as function of $\log N_{grp}$. Roughly 20\% of the central galaxies are host to radio sources even in the smallest systems and there is strong evidence for a rise in \rfrac\ with increasing $N_{grp}$. While this behavior is only at the one to two $\sigma$ level for \rfraca, the slope of the \rfrac-$\log N_{grp}$ relation increases for \rfracb\ and \rfracc. At the $3.5\sigma$ level for \rfracc, groups of 10 or more member galaxies are almost three times as likely to harbor a radio loud source in their central galaxy than smaller groups. Even though this trend may suggest even higher \rfrac\ in brightest cluster galaxies (BCGs), work by \citet{Best06a} determines a \rfrac\ maximum at approximately the value we find ($35\%$).

Two quantities used as an indicator of group mass are $\sigma_{v}$ and M$_{r,tot}$. While we are hindered  by low number statistics in the highest bins for $\sigma_v$, we see no statistically significant trend of \rfrac\ with the velocity dispersion of the group. There is marginal evidence for a correlation between \rfrac\ and M$_{r,tot}$, though not enough to be conclusive.

The central galaxy's properties may be more important than the group's for the radio fraction. To investigate, we look for any correlation between \rfrac\ and M$_{r,cen}$, (g-r)$_{cen}$, and $\Delta M_r$ (Figure~\ref{fig:clustering}). The luminosity of the central galaxy is the strongest indicator (at the $5\sigma$ level) of radio activity that we studied (Figure~\ref{fig:clustering}, upper right). When compared with the faintest central galaxies in our sample, \rfrac\ is increased in the most luminous galaxies by a factor of $\sim12$. This relationship had been shown previously by \citet{Best05a} who found that the radio-loud fraction was a strong function of host galaxy mass ($f_{radio-loud}\propto$ M$_{\star}^{2.5}$). If we assume that the mass to light ratios of the brightest group galaxies do not vary wildly, this work demonstrates the continuation of this trend towards lower mass central galaxies.

The color of the central galaxy (g-r)$_{cen}$ may also indicate the presence of a central radio source as this parameter does correlate with clustering properties. We find an increase at the $\sim 3\sigma$ level from $\sim20\%$ to $\sim35\%$ in \rfraca\ in the reddest central galaxies (Figure~\ref{fig:clustering}, lower left). The effect is even more pronounced when looking at the radio loud fraction via \rfracb\ and \rfracc\ and therefore redder, presumably older elliptical galaxies, have an increased probability of hosting a radio source. We do not find a significant AGN contribution to the spectra of our FRII sample and do not expect any systematic color enhancement due to AGN. Most of the central galaxies are quite red, yet it is interesting that the reddest galaxies are most prone to radio activity. \citet{Berlind06b} found that bluer galaxies were in more clustered environments, especially in the high mass regime. Their results combined with our own suggest that radio activity and clustering bias are not necessarily correlated.

Central galaxy identification with FoF group and cluster finding algorithms is not perfectly robust. \citet{Berlind06a} and the maxBCG method \citep{Koester07} both rely upon the observed correlation between galaxy luminosity and radial distance from the center of the potential well to designate central galaxies. The scatter in the luminosity-position relationship decreases with increasing luminosity gap between the brightest and next brightest group or cluster member ($\Delta M_r$).  We find a positive trend between $\Delta M_r$ and the radio fraction (Figure~\ref{fig:clustering}, bottom right). As $\Delta M_r$ increases from 0 to 1.75, radio fraction increases $\sim 3.5$, $7$, and $4$ fold (\rfraca, \rfracb, \rfracc, respectively). The luminosity dominance of a galaxy in the group environment, and hence its certainty as the central galaxy, is strongly correlated with radio emission. 

\section{Summary and Conclusions}\label{sec:con}

We present the first estimate of the maximum radio source lifetime in a well defined sample of groups and clusters of galaxies selected from the SDSS and FIRST surveys. Cross-correlating the group catalog identified by \citet{Berlind06a} with FIRST, we identify FRII sources associated with central group galaxies. The projected lengths of these sources are determined by the separation of the end of the lobe and either a core radio source or the optical position of the host galaxy. From this sample and measurement, we determined the most likely maximum lifetime of FRIIs through comparison with mock catalogs of known jet, lobe, and IGM properties. We modeled the linear growth of these mock sources with time and their luminosity evolution with the prescription set forth in KA and KDA. By fixing the jet, lobe, and environment properties, the maximum amount of time the jet is fed energetically by the AGN was determined via a maximum likelihood approach. Our prototype for the average state of FRII lobes and surrounding medium is listed in Table~\ref{table:LH} (\X{1}). We stress that our sample is representative of the group environment and as such our lifetime estimate is for a typical FRII source in these systems. While individual sources of lengths and lifetimes exceeding those seen here are observed, they are far less numerous than the members of this sample. Using the default input parameters to the KDA model and the Blundell $Q$ distribution, our sample's most likely \Tmax\ is $1.75(\pm^{0.14}_{0.07})\times 10^7$ years.

We numerically investigated the dependence of any single input parameter and \Tmax, while we present an analytic treatment in the Appendix. First, the power associated with the jets can affect the lifetime calculated in our analysis. We use two distributions: (1) One that reproduces the active radio galaxy luminosity function derived in \citet{Sadler02} and (2) The FRII jet power distribution found on energetic grounds from BRW. Keeping all other input variables fixed, we find that changing the jet power distribution from \Qsad\ to \Qblu\ produces only a $\sim 30\%$ change in the value of \Tmax\ ($1.22\times10^7$ to $1.75\times10^7$ yr). Mock radio sources were expanded into three distinct IGMs. In addition to the KDA density profile, the less dense profile of the BRW model and a recent observational measurement by \citet{Jetha07} were tested. As expected the calculated lifetime decreased relative to our fiducial \Tmax, but only marginally so: $\sim10\%$ for BRW and $\sim30\%$ for the Jetha profile. In practice, the density profile of the IGM in galaxy groups spans some distribution whose shape will contribute to a range of expansion speeds for FRII sources. The work here demonstrates that \Tmax\ is a weak function of environment and will likely vary by less than $50\%$. We find that the axial ratio of the lobe had the largest effect on the calculated lifetime of our observed sample. Still, the calculated lifetime of extreme axial ratio lobes was not different from our fiducial results by more than a factor of $\sim 2$. We sample the entire range of expected axial ratio and find \Tmax\ in the range of $\sim0.5-2.4\times 10^7$ years (\Qsad). Due to the sensitivity of FIRST and the relatively small distance to the groups in this study, uncertainties in the particle energy distribution do not effect our results. We created mock catalogs that only differed in the choice of the injection index $p$ and found \Tmax\ was statistically identical over the entire probable range. The values and associated uncertainties of parameters solely contributing to the luminosity of the sources are insignificant in this analysis.

Taken together, these tests on the parametric dependence of \Tmax\ demonstrate how robust our results are to ambiguity in the true distributions of many jet, lobe, and environmental parameters. Due to the self-similar nature of the length to age relationship determined in KA, these results would not change appreciably if another radio source evolution model was incorporated into our analysis. While individual extreme sources may be observed with ages significantly larger than the maximum lifetime found here, the vast majority of FRII sources reside within the parameter space examined. We therefore conclude that the average FRII lifetime is $1.5(\pm0.5)\times 10^7$ years.

This characteristic maximum lifetime of the FRII population is in good agreement with various methods of radio source dating for individual objects. Recently, O'Dea (2007, in prep.) used spectral aging to estimate the source ages of 31 powerful FRII radio galaxies. Under minimum energy conditions, all source ages were less than $1.1\times 10^7$ years while most were several Myr old. \citet{Allen06} presented a comprehensive study of luminous elliptical galaxies, measuring the age of (primarily FRI) radio sources still energetically associated with their parent AGN.  They estimate the cavity age using the sound speed of the IGM and the size of the bubbles in the X-ray gas and find ages of $10^6$ to $10^8$ years.

The maximum lifetime of $1.5 \times 10^7$ years for FRIIs is also in good agreement with observational constraints on the lifetime of QSOs, which suggest values from $10^6$ to $10^8$ years \citep[\eg][]{Martini04} and the value of the e-folding or Salpeter timescale for black hole growth at the Eddington rate $t_S = 4.5 \times 10^7$ yr (for 10\% radiative efficiency). This is most likely a coincidence, because these FRIIs show little evidence of nuclear activity in their SDSS spectra, are consequently accreting at a low rate, and also show no evidence that they have been triggered by mergers. In fact the detailed models of AGN triggered in major mergers find that the lifetime is progressively longer when the AGN is defined at a progressively lower luminosity or accretion rate limit \citep{Hopkins05}. At the accretion rates of these FRIIs, the predicted lifetime would be substantially longer than $10^7$ years in the merger scenario.

We investigated the connection between host galaxy and group properties and the probability of harboring a radio source. The size of a group correlates with \rfrac\ at the $2\sigma$ level. Assuming some average galactic mass, group size should be proportional to its viral radius. It is plausible that larger viral radii would encapsulate greater gas mass on average. If AGN feedback regulates star formation, larger systems would need shorter duty cycles or longer lifetimes of feedback activity to counteract their stronger cooling flows. The increase in \rfrac\ with group size supports the idea of episodic AGN heating.

The luminosity of the host galaxy is the strongest predictor of radio activity. Comparing galaxies with $M_r=-20.5 ... -23.0$, the radio fraction increased with luminosity by at least a factor of $5$ at the greater than $4\sigma$ level depending on the criterion used to determine the radio source activity threshold. This result is in agreement with previous studies of radio-loud AGN in BCGs \citep{Best06a, Croft07}. Our combined results point toward galaxy luminosity being the most important indicator of radio activity over a large range group/cluster mass. A slightly weaker correlation ($\sim3\sigma$) between host galaxy color and \rfrac\ was also found. Redder galaxies have an increased probability of harboring radio sources. Interestingly, galaxy color was found to be anti-correlated with clustering bias recently in the literature \citep{Berlind06b}. The two results suggest no connection between radio fraction and bias.

The luminosity gap between the brightest and second brightest group or cluster member has been discussed as a measure of certainty regarding central galaxy identification. The maxBCG catalog and that of \citet{Berlind06a} both designate the brightest cluster (group) member as the central galaxy. A caveat of this technique is uncertainty in the central galaxy's position with respect to the center of the group or cluster's potential well. As $\Delta M_r$ increases, the position-luminosity relationship tightens \citep{Loh06}. We find a significant correlation between \rfrac\ and $\Delta M_r$, which may be due in part to misdiagnosis of the central galaxy or that this galaxy is not at the center of the group potential. \citet{Best06a} examined the radio loud AGN fraction of BCG galaxies, finding an increased likelihood of radio source association when compared to field galaxies. The discrepancy in radio fraction between central and field galaxies may increase if the possible misclassification of central galaxies is taken into account. This potential source of incompleteness associated with all central galaxy studies should be addressed in future analysis.  

Our results have important implications for AGN feedback models. The recent literature includes many models for AGN energy injection into the ICM and its subsequent effects on galaxies and hot, gaseous halos \citep[\eg][]{Sijacki06}. Heating from AGN is now included in many structure formation models to match the observed galactic LF \citep[\eg][]{Croton06,Springel05a, Springel05b,Hopkins06}. The FRII lifetime presented here is an accurately measured timescale over which jets are energetically sustained by AGN and perform work on the ICM. Many models use an instantaneous injection of energy into the ICM as the timescales involved have historically had order of magnitude uncertainty. While it is true that FRIs are far more numerous than FRIIs, there is increasing evidence of FRIIs and FRIs being the manifestation of the \emph{same} underlying jet mechanism in different environments. A dramatic empirical demonstration of this are the hybrid morphology sources (HYMORS) in which one lobe of a double source appears to be an FRII while the other exhibits FRI characteristics \citep{Gopal-Krishna00, Gawronski06}. If FR morphology is a function of environment, it is reasonable to assume both classes share the same lifetime. We have shown the lifetime of FRII sources to be $\sim 10^7$ years in a wide range of initial conditions and it is most accurate to inject energy over this timescale.

\acknowledgements
We would like to thank Chris O'Dea for his helpful suggestions and Scott Gaudi for illuminating discussion. We also appreciate a thorough and helpful referee's report, which has significantly improved this paper. Funding for the Sloan Digital Sky Survey (SDSS) and SDSS-II has been provided by the Alfred P. Sloan Foundation, the Participating Institutions, the National Science Foundation, the U.S. Department of Energy, the National Aeronautics and Space Administration, the Japanese Monbukagakusho, and the Max Planck Society, and the Higher Education Funding Council for England. The SDSS Web site is http://www.sdss.org/. This research makes use of the FIRST and NVSS radio surveys.

\appendix{}
\section{Parameter Dependence}\label{sec:appendix}

Here we derive the analytic dependence of \Tmax\ on the other KDA model parameters. By plugging eq.~\ref{eq:tau} into eq.~\ref{eq:length} we obtain the full equation for the length of a lobe in terms of the parameters discussed in section~\ref{sec:KDA}:
\begin{equation}\label{eq:length2}
l = c(\beta, R_T) \left( \rho a^{\beta} \right)^{- 1/ ( 5 - \beta)} Q^{1/(5-\beta)} t^{3/ (5-\beta)},
\end{equation}
where $c(\beta, R_T)$ is just $c_1$ from eq.~\ref{eq:length} and eq.\ 4 (KDA), but shows that it is dependent upon $\beta$ and \Rt\ (see eq.\ 25 of KDA for the definition). We can easily manipulate this equation to solve for the age of the source, $t$. To zeroth order, we are trying to determine \Tmax\ such that a source at the average age $\bar{t}=t_{\rm max}/2$ has the median lobe length $\bar{l}$. Hence setting $t=\bar{t}$ and $l=\bar{l}$ we find
\begin{equation}\label{eq:time}
\bar{t} = \bar{l}^{(5-\beta) /3} c(\beta)^{(\beta - 5)/3} \left( \rho a^{\beta} \right)^{1/3} Q^{-1/3}.
\end{equation}
Note that $\bar{t}$ and therefore $t_{\rm max}$ only weakly depends on the density distribution, $\rho_0a_{0}^\beta$, and is independent of $p$. 

We first compare two different density distributions with $\rho_1$, $a_1$, $\beta_1$ and $\rho_2$, $a_2$, $\beta_2$, respectively. The distribution of jet powers is kept constant (either Sadler or BRW). In this case the two maximum ages are related by
\begin{equation}
\frac{t_{{\rm max},1}}{t_{{\rm max},2}} = \frac{\bar{t}_1}{\bar{t}_2} = \frac{\bar{l_1}^{(5-\beta_1 )/3}}{\bar{l_2}^{(5-\beta_2 )/3}} \frac{c(\beta_1)^{(\beta_1 - 5)/3}}{c(\beta_2)^{(\beta_2 - 5)/3}} \left( \frac{\rho_1 a_1^{\beta_1}}{\rho_2 a_2^{\beta_2}} \right)^{1/3}.
\end{equation}
Assigning the KDA density profile (\X{1}) to $t_{max,1}$; the density profile from \citet{Jetha07} (\X{3}) to $t_{max,2}$; and letting $\bar{l_1}=25.75$ kpc (\X{1S}) and  $\bar{l_2}=30.85$ kpc (\X{3S}), we find $t_{max, KDA, Q_S}$ / $t_{max, Jetha, Q_S} = 1.36$ and with a similar analysis $t_{max, KDA, Q_B}$ / $t_{max, Jetha, Q_B} =1.35$. These results are virtually identical to those found in our statistical approach, namely $t_{max, \X{1S}} / t_{max, \X{3S}}=  1.39$ and $t_{max, \X{1B}} / t_{max, \X{3B}}= 1.34$. Similarly, analytically comparing the KDA and BRW density profiles, $t_{max, KDA, Q_S}$ / $t_{max, BRW, Q_S} = 1.06$ and $t_{max, KDA, Q_B}$ / $t_{max, BRW, Q_B} = 1.10$ while $t_{max, \X{1S}} / t_{max, \X{2S}} = 1.06$ and $t_{max, \X{1B}} / t_{max, \X{2B}} = 1.10$. 

To isolate the effect on \Tmax\ by the choice of jet power distribution, we hold the density profile and axial ratio of the lobes fixed. In this case, the ratio of \Tmax\ simply becomes a function of the $Q$ distributions:
\begin{equation}\label{eq:Q_dep}
\frac{t_{{\rm max},sad}}{t_{{\rm max},blu}} = \frac{\bar{t}_{sad}}{\bar{t}_{blu}} =  \frac{\bar{l_1}^{(5-\beta_1 )/3}}{\bar{l_2}^{(5-\beta_2 )/3}} \left(\frac{Q_{sad}}{Q_{blu}} \right)^{-1/3}
\end{equation}

Taking the median value of each $Q$ distribution as a representative jet power and $\bar{l}$ from our mock catalogs, we find $t_{max, KDA, Q_S} / t_{max, KDA, Q_B} = 0.69$. Again, this analytical result is in excellent agreement with the simulation derived $t_{max, \X{1S}} / t_{max, \X{1B}}= 0.70$. Notice that the statistically obtained ratio of \Qsad\ to \Qblu\ across all parameter choices in table~\ref{table:LH} are between $0.67$ and $0.72$. A similar analysis for $\X{iS}/\X{iB}$ reveals that the theoretically and practically obtained ratios are in very good agreement. An increase in the median of the $Q$ distribution by a factor of $2.15$ (from \Qblu\ to \Qsad) decreases  \Tmax\ by only a factor of $\sim0.70$.

Finally, we examine the relationship between axial ratio and \Tmax. As $c_1$ and $\bar{l}$ are the only terms in equation~\ref{eq:time} that depend upon \Rt, the ratio of lifetimes derived from two models that only differ in their choice of \Rt\ is:
\begin{equation}\label{eq:Rt_dep}
\frac{t_{{\rm max},R_{T,1}}}{t_{{\rm max},R_{T,2}}} = \frac{\bar{t}_{R_{T,1}}}{\bar{t}_{R_{T,2}}} = \frac{\bar{l_1}^{(5-\beta_1 )/3}}{\bar{l_2}^{(5-\beta_2 )/3}} \frac{c(\beta_1)^{(\beta_1 - 5)/3}}{c(\beta_2)^{(\beta_2 - 5)/3}}.
\end{equation}
As an example, we find \Tmax(\Rt$=2.0$,\Qsad) / \Tmax(\Rt$=6.0$,\Qsad) $= 2.33$. This expectation is met very well as the ratio of \X{1S} / \X{6S} $=2.30$. In fact, eq.~\ref{eq:Rt_dep} is a good fit to the \Tmax\ distribution resulting from simulations that vary in \Rt.

The equations here derive only from the self similar solution for the length of these FRII sources (eq.~\ref{eq:length}). This particular equation is a characteristic of all the radio source evolution models discussed in this paper (KA, KDA, BRW, MK). We have shown that \Tmax\ is a weakly varying function of the density profile, jet power distribution, and axial ratio. The true nature of these jets, lobes, and their environments must be systematically and significantly different from the assumptions adopted here to change our results by a factor of 3 or higher.


\end{document}